\documentclass[preprint]{emulateapj}
\usepackage{latexsym}
\usepackage{amssymb}
\usepackage{amsmath}
\usepackage{epsfig}
\usepackage{appendix}
\usepackage{color,ulem}

\singlespace

\def\msun{\,{\rm M_\odot}}
\def\smsun{\,{\rm M_\odot \,kpc^{-2}}}
\def\lsun{\,{\rm L_\odot}}
\def\kms{\,{\rm km \, s^{-1}}}

\def\spose#1{\hbox to 0pt{#1\hss}}
\def\lta{\mathrel{\spose{\lower 3pt\hbox{$\mathchar"218$}}
     \raise 2.0pt\hbox{$\mathchar"13C$}}}
\def\gta{\mathrel{\spose{\lower 3pt\hbox{$\mathchar"218$}}
     \raise 2.0pt\hbox{$\mathchar"13E$}}}

\lefthead{Rashkov et al.}
\righthead{On the assembly of the Milky Way dwarf satellites}
\submitted{Accepted for publication in The Astrophysical Journal}


\begin{document}
\title{On the assembly of the Milky Way dwarf satellites and their common mass scale}

\author{Valery Rashkov$^{1}$, Piero Madau$^{1}$, Michael Kuhlen$^{2}$, \& J\"urg Diemand$^{3}$}
 
\altaffiltext{1}{Department of Astronomy and Astrophysics, University of California,  1156 High Street, Santa Cruz, CA 95064.} 
\altaffiltext{2}{Theoretical Astrophysics Center, University of California Berkeley, 601 Campbell Hall, Berkeley, CA 94720.}
\altaffiltext{3}{Institute of Theoretical Physics, University of Z\"urich, Winterhurerstrasse 190, CH-8057 Z\"urich, Switzerland.} 

\begin{abstract}
We use a particle tagging technique to dynamically populate the $N$-body {\it Via Lactea II} high-resolution simulation with stars. The method is calibrated using the observed luminosity function of Milky Way satellites and the concentration of their stellar populations, and self-consistently follows the accretion and disruption of progenitor dwarfs and the build-up of the stellar halo in a cosmological ``live host". Simple prescriptions for assigning stellar populations to collisionless particles are able to reproduce many properties of the observed Milky Way halo and its surviving dwarf satellites, like velocity dispersions, sizes, brightness profiles, metallicities, and spatial distribution. Our model predicts the existence of approximately 1,850 subhalos harboring ``extremely faint" satellites (with mass-to-light ratios $>5 \times 10^3$) lying beyond the {\it Sloan Digital Sky Survey} detection threshold. Of these, about 20 are ``first galaxies", i.e. satellites that formed a stellar mass above $10\,\msun$ before redshift 9. The ten most luminous satellites ($L>10^6\,\lsun$) in the simulation are hosted by subhalos with peak circular velocities today in the range $V_{\rm max}=10-40\,\kms$ 
that have shed between 80\% and 99\% of their dark mass after being accreted at redshifts $1.7<z<4.6$. The satellite maximum circular velocity $V_{\rm max}$ and stellar line-of-sight velocity dispersion $\sigma_{\rm los}$ today follow the relation $V_{\rm max}=2.2\sigma_{\rm los}$. We apply a standard mass estimation algorithm based on Jeans modelling of the line-of-sight velocity dispersion profiles to the simulated dwarf spheroidals, and test the accuracy of this technique. The inner (within 300 pc) mass-luminosity relation for currently detectable satellites is nearly flat in our model, in qualitative 
agreement with the ``common mass scale" found in Milky Way dwarfs. We do, however, predict a weak, but significant positive correlation for these objects: $M_{300}
\propto L^{0.088 \pm 0.024}$. 
\end{abstract}

\keywords{Galaxy: halo -- kinematics and dynamics -- galaxies: dwarf -- formation -- structure -- methods: numerical}
 
\section{Introduction}

Galaxy halos are one of the crucial testing grounds for $\Lambda$CDM structure formation scenarios, as they contain the imprints of their assembly via the hierarchical merging and accretion of many smaller progenitors \citep{whi78,blu84}. Subunits are predicted to collapse at high redshift and have high density concentrations, and when they merge into more massive hosts they are able to resist tidal disruptions and survive in large numbers as bound substructures \citep[e.g.][]{die08}. The dwarf spheroidal (dSph) satellites present today in the halos of massive galaxies like our own Milky Way (MW) are thought to represent a fossil record of such early merging activity. The old tally of MW ``classical" dwarfs with $M_V \lesssim -10$ \citep[e.g.][]{mat98} has been recently 
revised by the discovery of a new, large population of ultra-faint dwarfs in the {\it Sloan Digital Sky Survey} (SDSS) \citep{wil05,bel07}.
Follow-up spectroscopic observations have shown that these extreme objects are the most dark matter-dominated galaxies as well as some of the most metal-poor stellar systems known in the universe \citep{sim07,kir08,geh09}. While most of the observed dSphs seem to be in pressure-supported equilibrium, others (like the Sagittarius dwarf, see \citealt{iba01,maj03}) are being ripped by tidal forces into long streams of stars that will eventually dissolve into the Galactic stellar halo. Large coherent stellar streams have been observed in Andromeda's halo \citep{mcc09} and in the outskirts of other nearby galaxies \citep{mar10,mou10}.

The predicted counts of bound substructures vastly exceed the number of known satellites of the MW, a ``missing satellite problem'' that has been the subject of many numerical studies over the last decade \citep[e.g.][]{kly99,moo99,mad08a}. While this discrepancy is partially alleviated by correcting the data to account for the incomplete sky coverage and distance-magnitude or surface brightness limits of current surveys \citep{kop08,tol08,bul10}, it is generally agreed that a complete solution to the apparent conflict between the Galaxy's relatively few observed stellar clumps and the abundance of halo substructure in cold dark matter simulations requires either a modification of the nature of the dark matter particle that would act to suppress small-scale power \citep[e.g.][]{col00} or an extremely low star formation efficiency in small dark matter subunits \citep[e.g.][]{kau93,ben02,bul00,som02,kra04,mad08b,kuh11}.

Measurements of the line-of-sight velocity dispersion as a function of radius in dwarf satellites, combined with a spherical Jeans analysis, may have provided another ``small-scale puzzle". Despite spanning about 5 decades in luminosity, MW dSphs appear to contain a similar amount of dark mass, $M_{300}\sim 10^7\,\msun$, within a fixed aperture radius of 300 pc from their centers \citep{str08,wal09}.
It is unclear at this stage whether the flat inner mass-luminosity relation inferred, $M_{300}\propto L^{0.03\pm 0.03}$, is an accident of small number statistics, an artifact of the spherical Jeans analysis technique, arises from an observational selection bias \citep{bul10}, is evidence for 
a new fundamental scale in galaxy formation \citep{mun09}, or arises naturally in the $\Lambda$CDM paradigm \citep[e.g.][]{mac09,stri10}.

Numerical simulations of the assembly of the halo of Milky Way-sized galaxies have the potential to shed light on the nature of dSph galaxies and test the $\Lambda$CDM hierarchical clustering paradigm. Owing to their computational expense, state-of-the-art cosmological hydrodynamic 
simulations that include gas cooling, star formation, and supernova feedback, and can self-consistently follow the formation and evolution of MW substructure down to the present epoch, are limited to dark particle masses $m_p\gta 10^5\,\msun$ \citep{gue11}, and can well
resolve only the most massive subhalos. To be able to reach the necessary resolution, we use in this paper a particle tagging technique to dynamically populate the $N$-body {\it Via Lactea II} (hereafter VLII) high-resolution simulation with stars, and self-consistently follow the accretion 
and disruption of progenitor dwarfs and the build-up of the stellar halo in a cosmological ``live host". We show how simple prescriptions for assigning stellar populations to collisionless particles are able to reproduce many properties of the observed MW halo and its surviving dwarf satellites, like velocity dispersions, sizes, brightness profiles, metallicities, and spatial distribution. 
More importantly, with over 1.6 million stellar particles tagged in post-processing, we can then mimic the observations of the line-of-sight velocity dispersion profiles of the simulated dSphs, apply a standard mass estimation algorithm based on spherical Jeans modelling, test the accuracy of this technique, and confirm or dispel the existence of a common mass scale for satellite galaxies.  

A number of previous implementations of the particle-tagging approach exists in the literature \citep[e.g.][]{whi00,die05,bul05,moo06,coo10,tum10}. The approximation breaks down when baryons become dynamically important, i.e. in the inner regions of the MW and in the galactic disk. The orbits of some subhalos will pass through these regions, and one has to keep in mind and eventually correct for the fact that some substructure will actually feel the potential of baryonic configurations. Dwarf spheroidal satellites on the other hand seem to be dark matter-dominated on all scales, and our simulated subhalo potentials can be directly compared to the observed stellar kinematics.
VLII's high mass resolution allows the survival of small dark matter subhalos long after they are accreted by the main host. Resolved at the present epoch with anywhere between 1,000 and 100,000 particles, these systems, in combination with an appropriate technique to designate stellar particles, serve as a testbed for the techniques commonly used to estimate the masses of dSph galaxies from real data.

The paper is organized as follows. In \S~2 we describe the VLII simulation and particle tagging technique, together with the resulting stellar halo and the population of 
dwarf satellites surviving today. In \S~3 we describe the Jeans analysis used to estimate masses in pressure-supported collisionless systems, and assess its performance in correctly estimating the underlying mass distributions of the simulated dSph galaxies. We summarize our findings in \S~4.

\begin{table}
\centering
\caption{Build-up of the VLII stellar halo}
\begin{tabular}{l l l l}
\hline \hline
Snapshot & Redshift & $N_{\rm halos}$ & $M_{*} [\msun]$\\ 
\hline
0 & 27.54 & 0 & -\\ 
1 & 23.13 & 0 & -\\ 
2 & 17.88 & 0 & -\\ 
3 & 14.78 & 0 & -\\ 
4 & 12.71 & 3 & $4.48 \times 10^0$\\ 
5 & 11.20 & 24 & $9.83 \times 10^3$\\ 
6 & 9.14 & 190 & $1.89 \times 10^5$\\ 
7 & 7.77 & 530 & $4.25 \times 10^8$\\ 
8 & 4.56 & 1109 & $9.61 \times 10^7$\\ 
9 & 3.24 & 609 & $1.26 \times 10^8$\\ 
10 & 2.50 & 303 & $7.12 \times 10^7$\\ 
11 & 2.00 & 166 & $8.86 \times 10^8$\\ 
12 & 1.65 & 67 & $4.50 \times 10^7$\\ 
13 & 1.37 & 61 & $3.63 \times 10^3$\\ 
14 & 1.16 & 45 & $6.86 \times 10^5$\\ 
15 & 0.98 & 22 & $4.45 \times 10^5$\\ 
16 & 0.83 & 17 & $1.50 \times 10^7$\\ 
17 & 0.70 & 12 & $9.84 \times 10^4$\\ 
18 & 0.58 & 15 & $2.01 \times 10^4$\\ 
19 & 0.48 & 14 & $6.07 \times 10^3$\\ 
20 & 0.39 & 6 & $1.18 \times 10^1$\\ 
21 & 0.31 & 3 & $9.20 \times 10^1$\\ 
22 & 0.24 & 5 & $3.68 \times 10^2$\\ 
23 & 0.17 & 1 & $4.81 \times 10^2$\\ 
24 & 0.11 & 1 & $3.66 \times 10^0$\\ 
25 & 0.05 & 1 & $1.04 \times 10^1$\\ 
26 & 0.00 & 0 & -\\
\hline
\end{tabular}
\tablecomments{Columns 1 and 2 give the VLII snapshot number and its corresponding redshift. Columns 3 and 4 list the number of subhalos that reached their maximum dark matter mass $M_{\rm h}$ in that snapshot, together with the total stellar mass they contribute to the VLII main host at $z=0$.} 
\label{table:snap}
\vspace{+0.3cm}
\end{table}

\begin{figure*}
\centering
\includegraphics[width=0.49\textwidth]{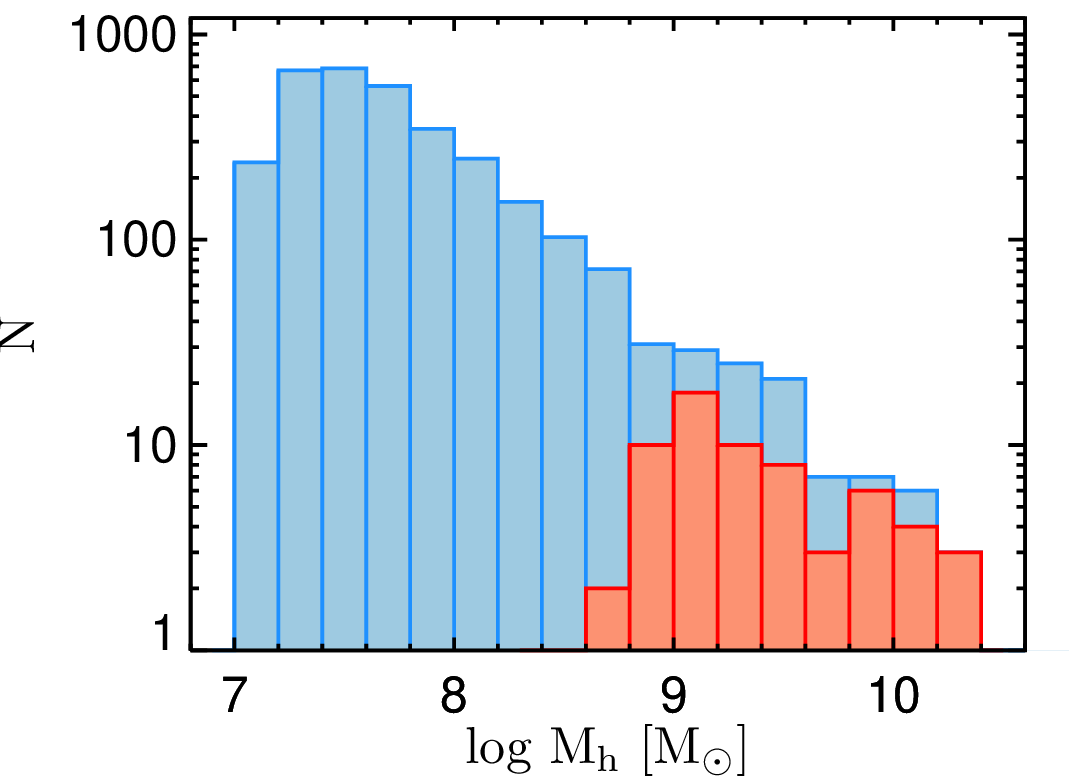}
\includegraphics[width=0.49\textwidth]{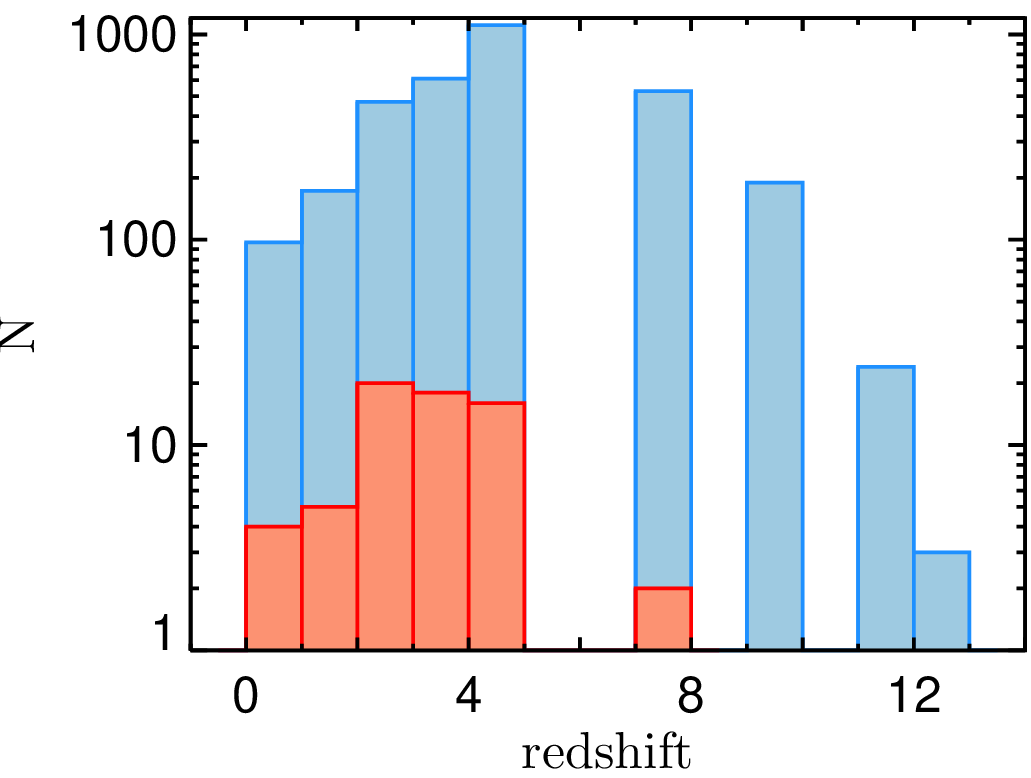}
\caption{Distributions of the peak mass $M_{\rm h}$ at infall for 3,204 VLII subhalos (in blue) above $M_{\rm h}=10^7\,\msun$ ({\it left panel}) and of their infall redshift ({\it right panel}). Tagged subsets of dark matter particles in these systems are followed in post-processing from infall to the present. Subhalos that at $z=0$ host a SDSS-detectable dSph populate the higher-mass and lower-redshift ends of these distrubutions (in red). The gaps in the redshift distribution correspond to $\Delta z = 1$ intervals where VLII snapshots were not analyzed in detail (see Table \ref{table:snap}).
}
\label{fig1}
\vspace{+0.3cm}
\end{figure*}

\section{Simulation and particle tagging technique} 
\label{sec:simmodel} 

The VLII simulation, one of the highest-precision calculations of the assembly of a Galactic halo to date \citep{die08,zem09}, was performed with the PKDGRAV tree-code \citep{sta01}. Initial conditions were based on the WMAP 3-year data release cosmology \citep{spe07} and were generated with a modified, parallel version of the GRAFIC2 package \citep{ber01}. VLII employs just over one billion $4,100\,\msun$ particles to model the formation of a $M_{200}=1.9\times 10^{12}\,\msun$ Milky-Way size halo and its substructure. About 40,000 subhalos of masses as low as $10^6\,\msun$ are resolved today within the host's $r_{200}=402$ kpc (the radius enclosing an average density 200 times the mean matter value).

The particle tagging technique makes full use of 27 snapshots (out of 400 total) of the simulation, spanning the assembly history of the host between $z=27.54$ and the present. Numbered from 0 to 26, the snapshots are listed in Table \ref{table:snap} together with the corresponding redshift. In each snapshot, all (sub)halos were identified with the 6DFOF group finder \citep{die06}, and linked from snapshot to snapshot to their most massive progenitor:  the halo `tracks' built in this way contain all the time-dependent structural information necessary for our study.
We use them to identify the simulation snapshot in which each subhalo reaches its maximum mass $M_{\rm h}$ before being accreted by the main host and stripped. By neglecting all subhalos with $M_{\rm h}<10^7\,\msun$, we restrict our analysis to 3,204 such tracks that contribute mass to the main host halo at $z=0$.
The mass distribution of the selected subhalos at infall and the distribution of infall redshifts are shown in Figure~\ref{fig1}. The gaps in the redshift histogram correspond to epochs in which no VLII snapshots were studied in detail (see Table \ref{table:snap}). Due to the great computational effort necessary in the analysis of a single VLII snapshot, 6DFOF was applied only to the 27 ones listed here, and expanding this analysis to more snapshots is beyond the scope of our work. Halos that are accreted by the main host in these intervals are tagged at the next available snapshot. As the halos we consider are composed of a large number of particles, they are well resolved even $\sim$0.5-0.8 Gyr (typical snapshot separation) after accretion, making it unlikely that this time discretization affects significantly our results.

A subset of dark matter particles in each subhalo is then ``tagged" as stars at infall, and no subsequent star formation is assumed to take place - hydrodynamical simulations suggest that ram pressure stripping would deplete the subhalo gas reservoir and shut star formation off shortly thereafter \citep[e.g.][]{may06}. It is therefore at infall that tagged particles acquire a stellar mass $m_{\rm sp}$, an age $t_{\rm sp}$, and a metallicity [Fe/H]$_{\rm sp}$: any subsequent evolution of the tagged stellar populations is purely photometric and kinematical in character, as systems age and are accreted and disrupted in a cosmological ``live host".

The following three prescriptions determine $m_{\rm sp}$, $t_{\rm sp}$, and [Fe/H]$_{\rm sp}$, and were tuned to match certain structural and statistical properties of the surviving dSph population at $z=0$: the star formation efficiency $M_* / M_{\rm h}$ is fit to reproduce the satellite luminosity function, the selection of particles painted in halos is  made to fit the current extent of their luminous component, and the assigned metallicities are designed to place satellites on the observed metallicity--luminosity relation today. The first two prescriptions thus determine the value $m_{\rm sp}$ for each tagged particle, while the third determines its [Fe/H]$_{\rm sp}$. The age $t_{\rm sp}$ of the stellar population consistently follows from the star formation efficiency prescription and the history of dark matter build-up of the subhalo it was tagged in.

In the spirit of \citet{kop09} and \citet{kra10}, the total stellar mass tagged at infall is assumed to scale as a power-law of the host subhalo mass,  
\begin{equation}
{\frac{M_*}{M_{\rm h}}}= 1.6\times 10^{-5}\, \left({\frac{M_{\rm h}}{10^9\,\msun}}\right)^{1.8}.
\label{eq:mstar}
\end{equation}
\citet{kra10} used a slope of 1.5 but argued that the results would not change drastically if a somewhat steeper (2.0-2.5) slope was assumed. \citet{kop09} similarly adopted a slope of 2.0. Our relation has a somewhat higher normalization than both. The difference likely arises from the conversion of the assigned stellar mass to a V-band luminosity. \citet{kop09} assume a mass-to-light ratio $M/L=1$ to be typical for old stellar populations, while we find that $M/L \sim 3-4$ better fits the range of ages and metallicities we assign to the infalling satellites (see below). A similar argument holds for the \citet{kra10} relation, which directly assigns a V-band luminosity to a halo of a given mass, instead of a stellar mass. Additionally, some of the halos in VLII lose a fraction of their stellar mass after accretion from tidal stripping, an effect that may not have been treated in the other two studies.

Our stellar mass-subhalo mass relation at infall has a scaling similar to the low-mass end ($M_h\sim 10^{11.5}\,\msun$) galaxy stellar mass--halo virial mass relation, $M_*/M_h\propto M_h^{1.9}$, derived at $z=1$ by \citet{behroozi10} using an abundance matching technique. 
It implies a star formation efficiency (the fraction of baryons turned into stars), $f_*\equiv (\Omega_M/\Omega_b)(M_*/M_{\rm h})$, that is close to 1\% in $M_{\rm h}=10^{10}\,\msun$ subhalos, and becomes negligibly small in systems below $M_{\rm h}=10^8\,\msun$. 
The detailed physical explanation of this behavior is still an area of active research. Here we stress that it is the simple prescription in equation (\ref{eq:mstar}) that largely determines the luminosity function of surviving dwarf satellites today and empirically bypasses the missing satellite problem by strongly suppressing star formation in low-mass systems.

The total stellar mass within an infalling subhalo is equally distributed to its 1\% most tightly bound dark matter particles. This number determines the concentration of the stellar system at the time of infall, and therefore governs the amount of stellar material stripped at later times as well as the present-day structural properties of surviving satellites, like their half-light radii and central surface brightnesses. 
\citet{coo10} have shown that tagging the 1\% highest total binding energy particles with stellar populations provides a good fit to the distribution of half-light radii in MW dSphs. Our analysis below confirms this result.

The metallicity assigned to each particle at infall is also a function of $M_{\rm h}$,
\begin{equation}
{\rm [Fe/H]}_{\rm sp} = -7.87 + 0.9 \times \log \left(\frac{M_{\rm h}}{10^3\,\msun}\right), 
\label{eq:met.lum}
\end{equation}
i.e. more massive subhalos are assumed to retain more of their enriched material and have a higher [Fe/H] than smaller systems. All particles initially tagged in the same subhalo are assigned the same metallicity. This prescription does not account for the timescale of enrichment of the stellar systems with iron or alpha-elements. In the present work metallicity only enters in the calculation of the luminosity of a given stellar system, and does not affect our kinematic studies. Two processes turn the above scaling into the observed present-day luminosity-metallicity relation of \citet{kir08}: (i) stellar mass loss from tidal stripping and (ii) dimming of the stellar populations with age (see the right panel of Fig.~\ref{fig3}). When comparing the above relation to the one derived for SDSS galaxies, we note that it lies below the extrapolation of the \citet{tre04} relation. This suggests that processes governing metal formation and retention in dSphs may be significantly different from those in more massive galaxies.

Finally, while all particles in a given subhalo are tagged at a single epoch (infall), they have a distribution of stellar ages resulting from applying equation (\ref{eq:mstar}) to the history of dark matter build-up of each halo prior to its infall and assigning the respective age $t_{\rm sp}$ to the corresponding fraction of tagged stellar mass. This process mimics the extended star formation histories of dwarf satellites in the MW halo \citep[see][]{orb08}.

After all stellar particles are tagged with the above properties, they are traced down to the $z=0$ snapshot in the simulation. Some are still part of surviving dSph satellites; many form coherent stellar streams and contribute to the smooth stellar halo component. 
Their final V-band luminosities are self-consistently determined for their age, stellar mass,
and metallicity from the stellar population synthesis models of \citet{bru03} (assuming a Chabrier initial mass function, \citealt{cha03}).

\subsection{Assembly history}
\label{sec:halo}

A total of 1.6 million tagged particles end up in the main host at redshift $z=0$ (0.35\% of all particles within $r_{200}$): most are tidally stripped from their subhalos after infall, and only 52\% of them remain bound to the 1,925 self-gravitating clumps that partially survive the accretion event.  The total mass of the stellar halo today is ${\cal M}_*=1.6 \times 10^9\,\msun$ within 402 kpc (of which $2.3\times 10^8\,\msun$ is in bound substructures), and Table \ref{table:snap} lists the contribution to ${\cal M}_*$ from each snapshot: the peak contribution comes from redshift 2. The mass in the stripped component between 1 kpc and 40 kpc is $5 \times 10^8\,\msun$, comparable to recent estimates for the MW halo mass within the same range \citep{bel08}.

In terms of its mass assembly history, the stellar component of the VLII halo is already in place by redshift $z=1.5$, while the dark component continues to grow from mergers with smaller subunits down to the present epoch. The faster assembly of the stellar component follows from the very steep mass-star formation efficiency relation adopted in our model. A similar behavior was seen in the assembly history of the six Aquarius stellar halos by \citet{coo10}. About 50\% of today's stellar halo can be traced back to a subhalo of $M_*=8.8 \times 10^8\,\msun$ that fell in at $z=2$. The three next most massive contributors, with $M_*/\msun=(4.2 \times 10^8, 1.2 \times 10^8, 4.5 \times 10^7)$, were accreted at redshifts $z=$7.8, 3.2, and 1.6, respectively. Combined, these four progenitor systems account for about 88\% of the total stellar halo. 

\begin{figure*}
\begin{center}
$\begin{array}{c@{\hspace{1mm}}c@{\hspace{1mm}}c}

\includegraphics[width=59mm]{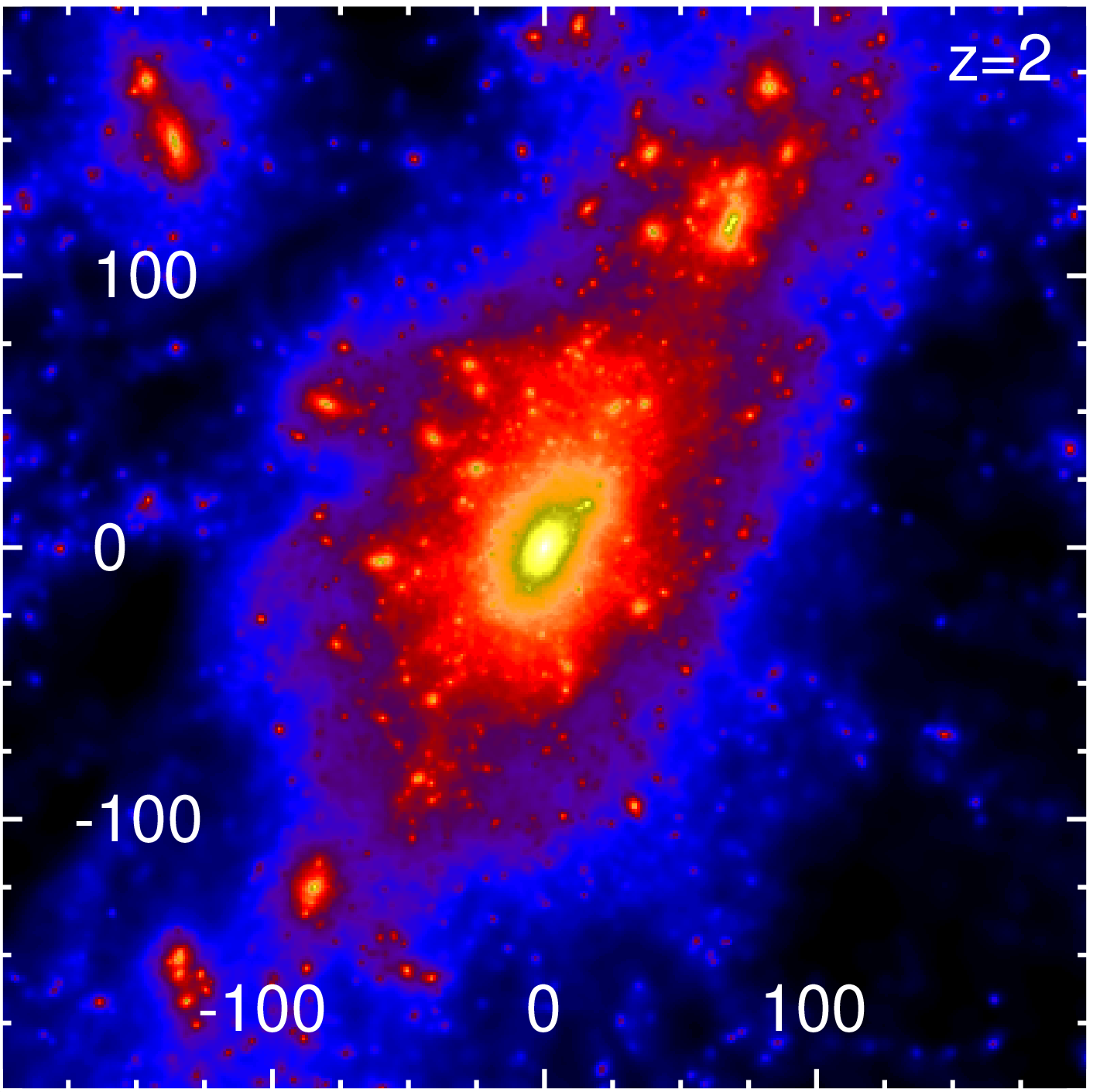} &
\includegraphics[width=59mm]{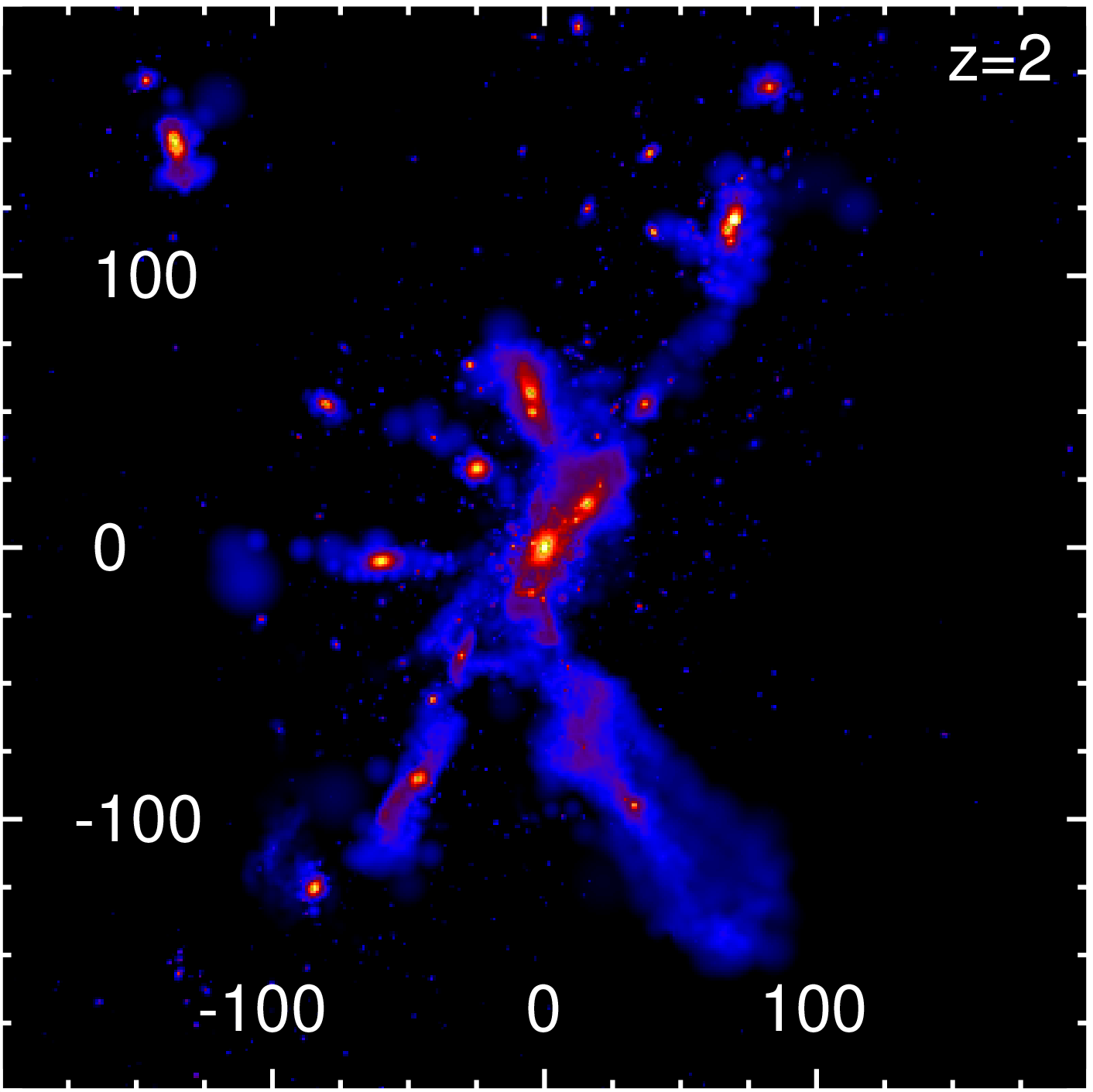} &
\includegraphics[width=59mm]{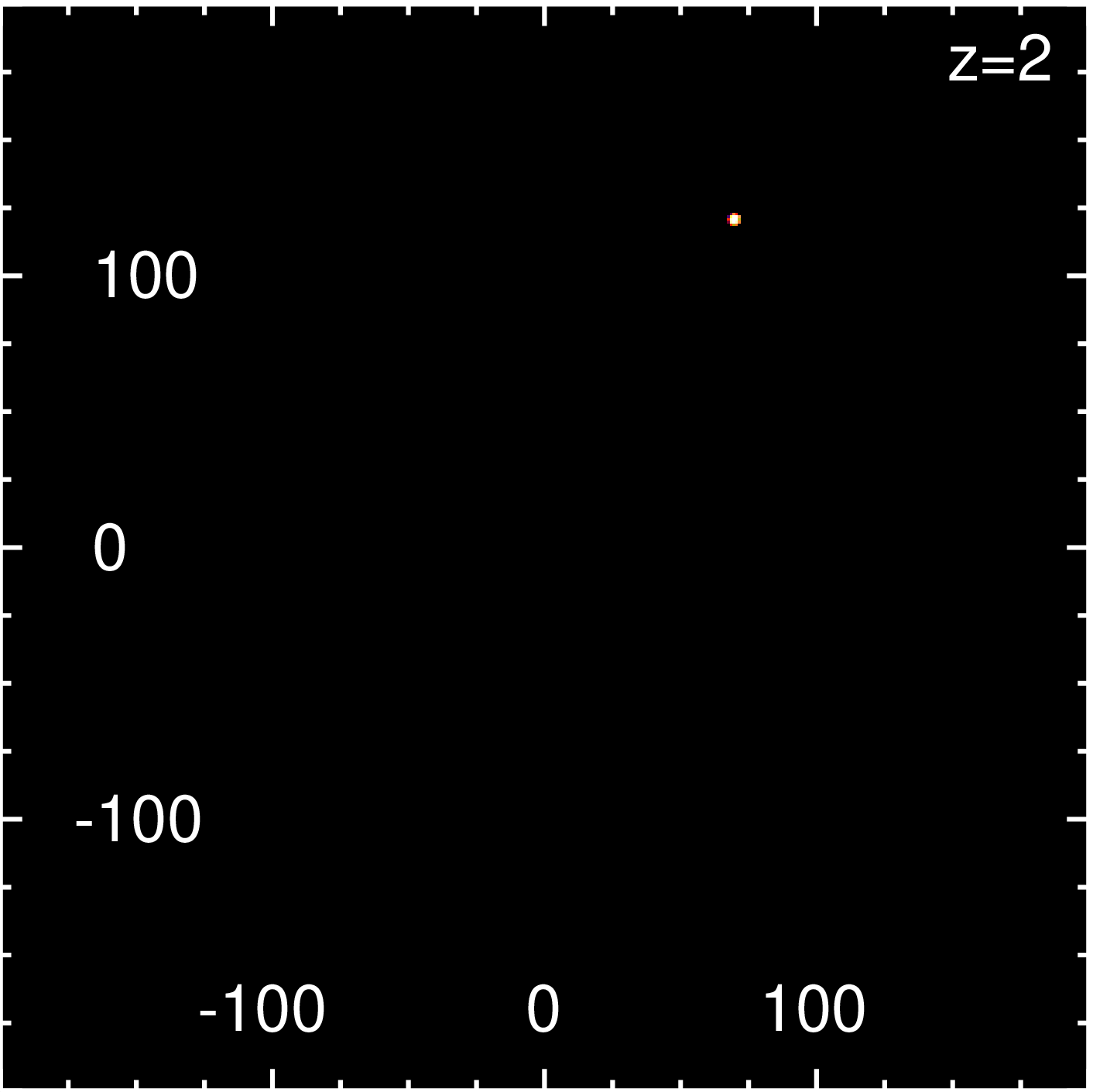} \\
\includegraphics[width=59mm]{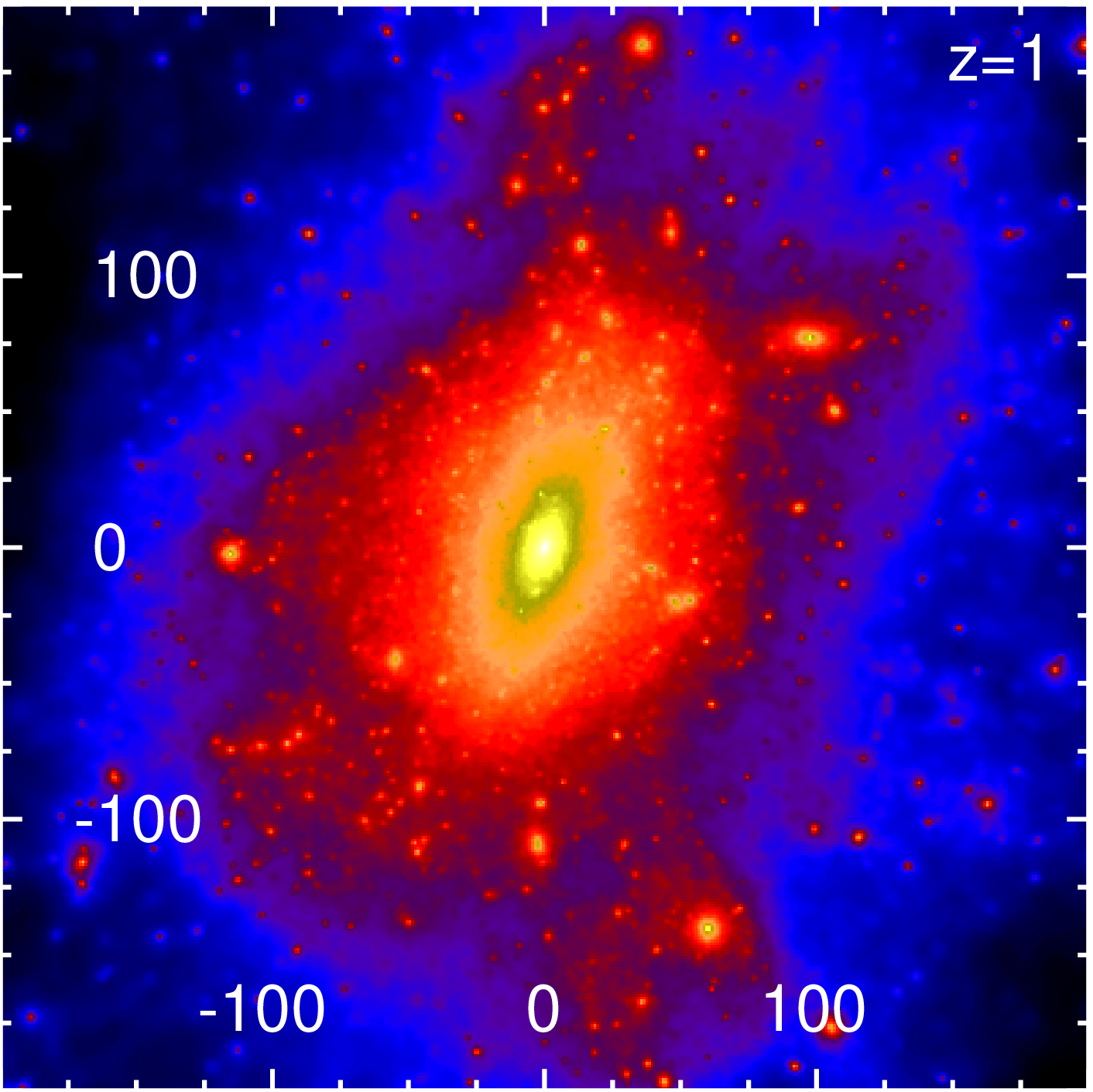} &
\includegraphics[width=59mm]{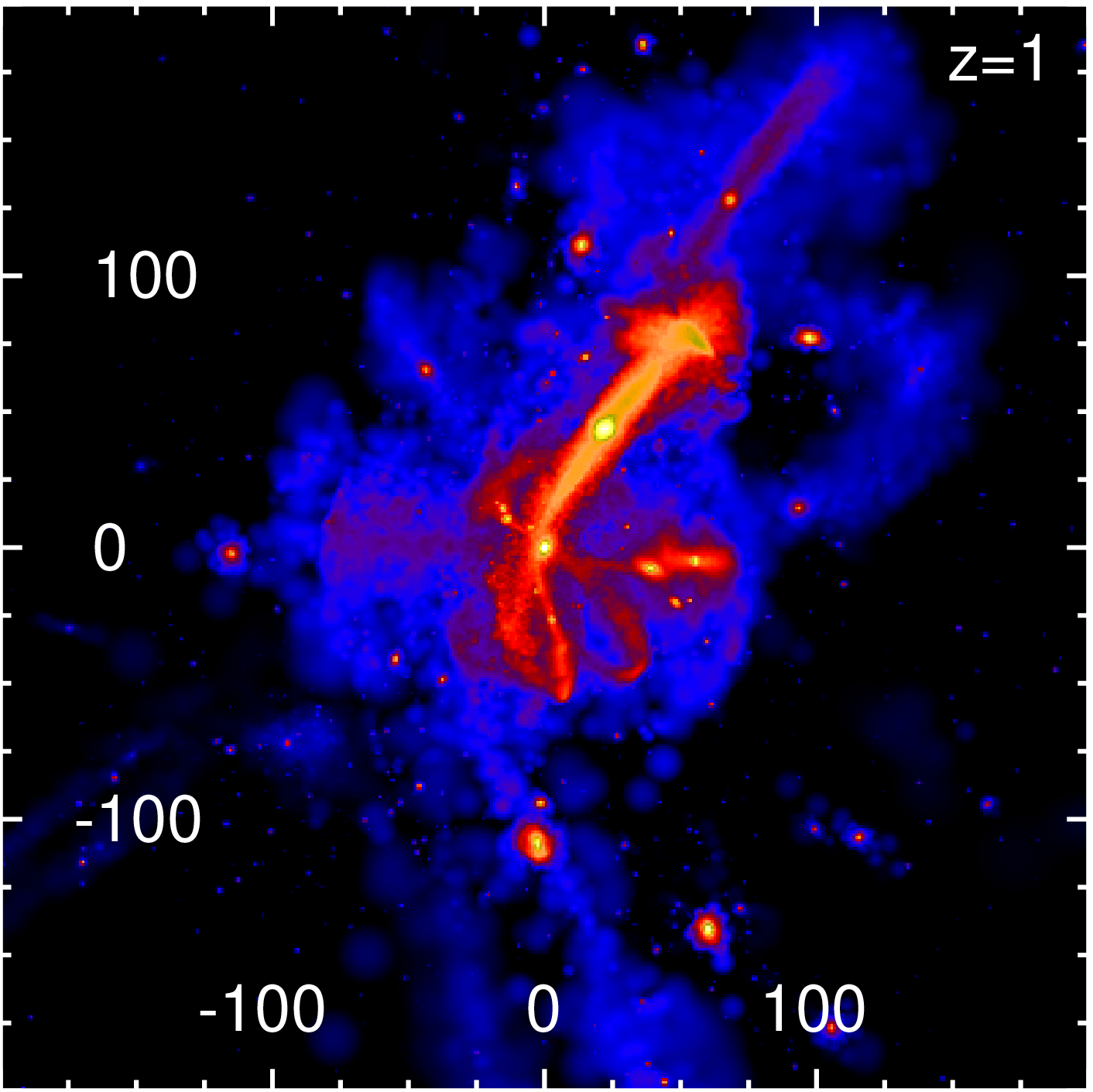} &
\includegraphics[width=59mm]{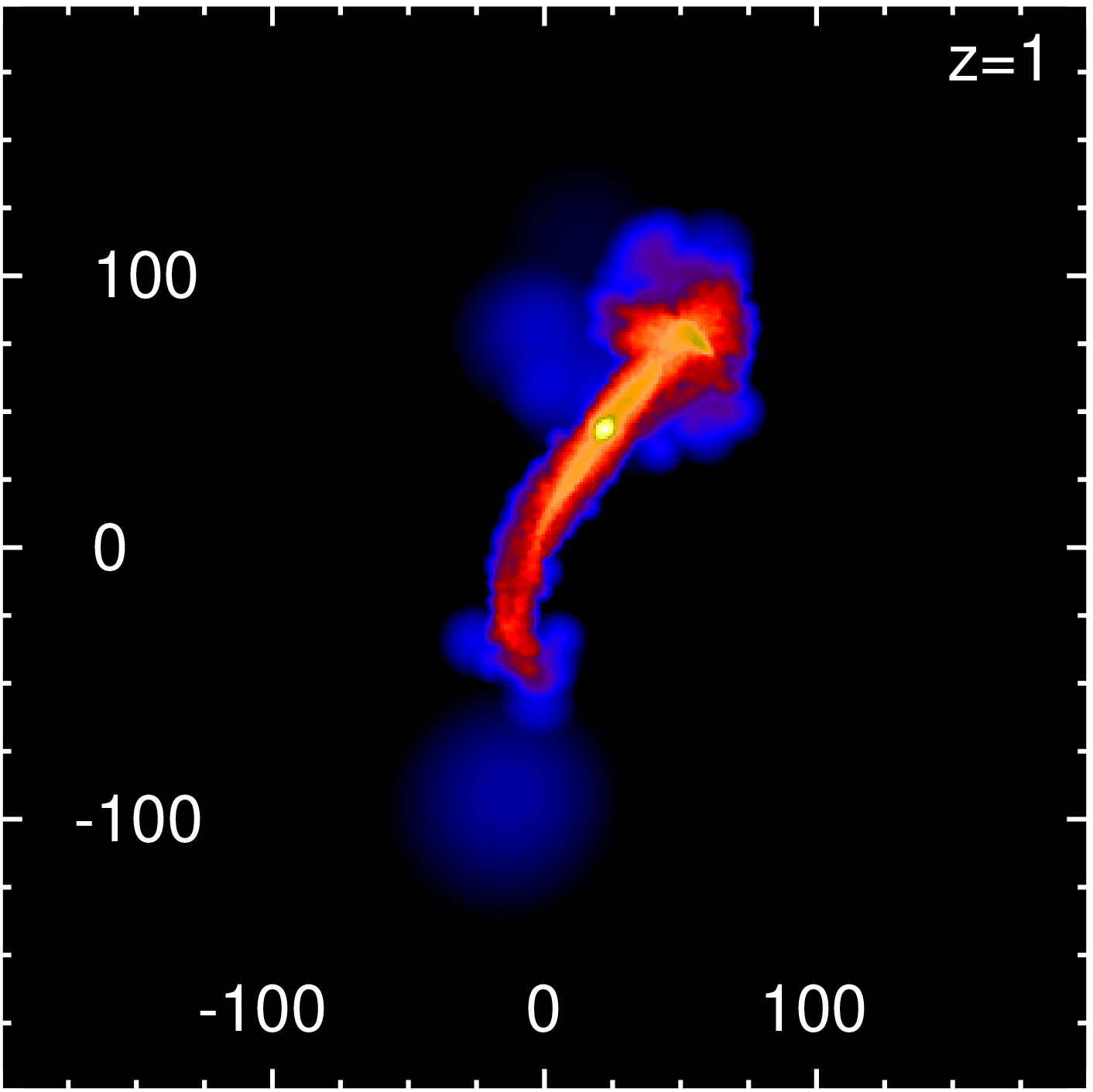} \\
\includegraphics[width=59mm]{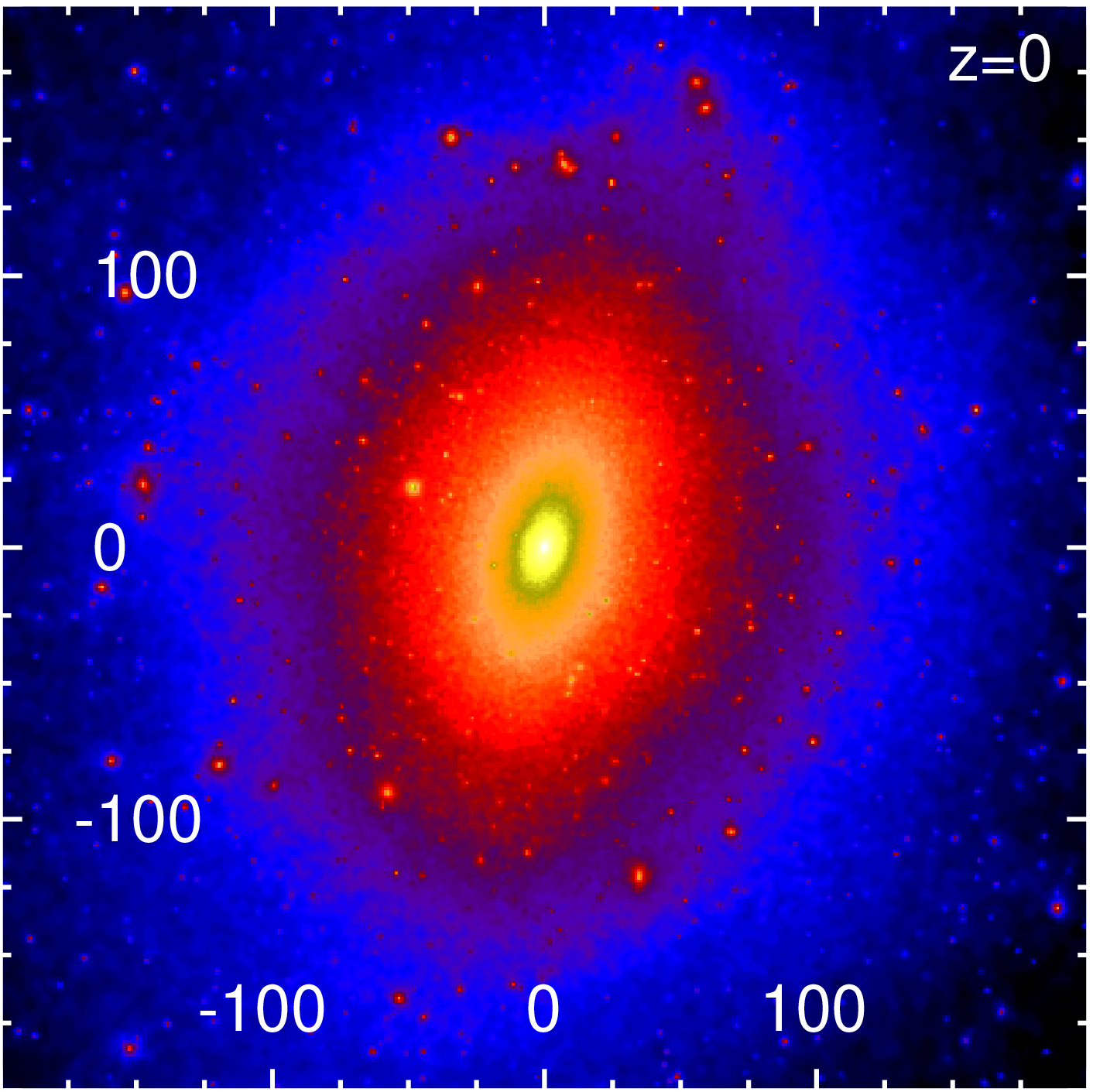} &
\includegraphics[width=59mm]{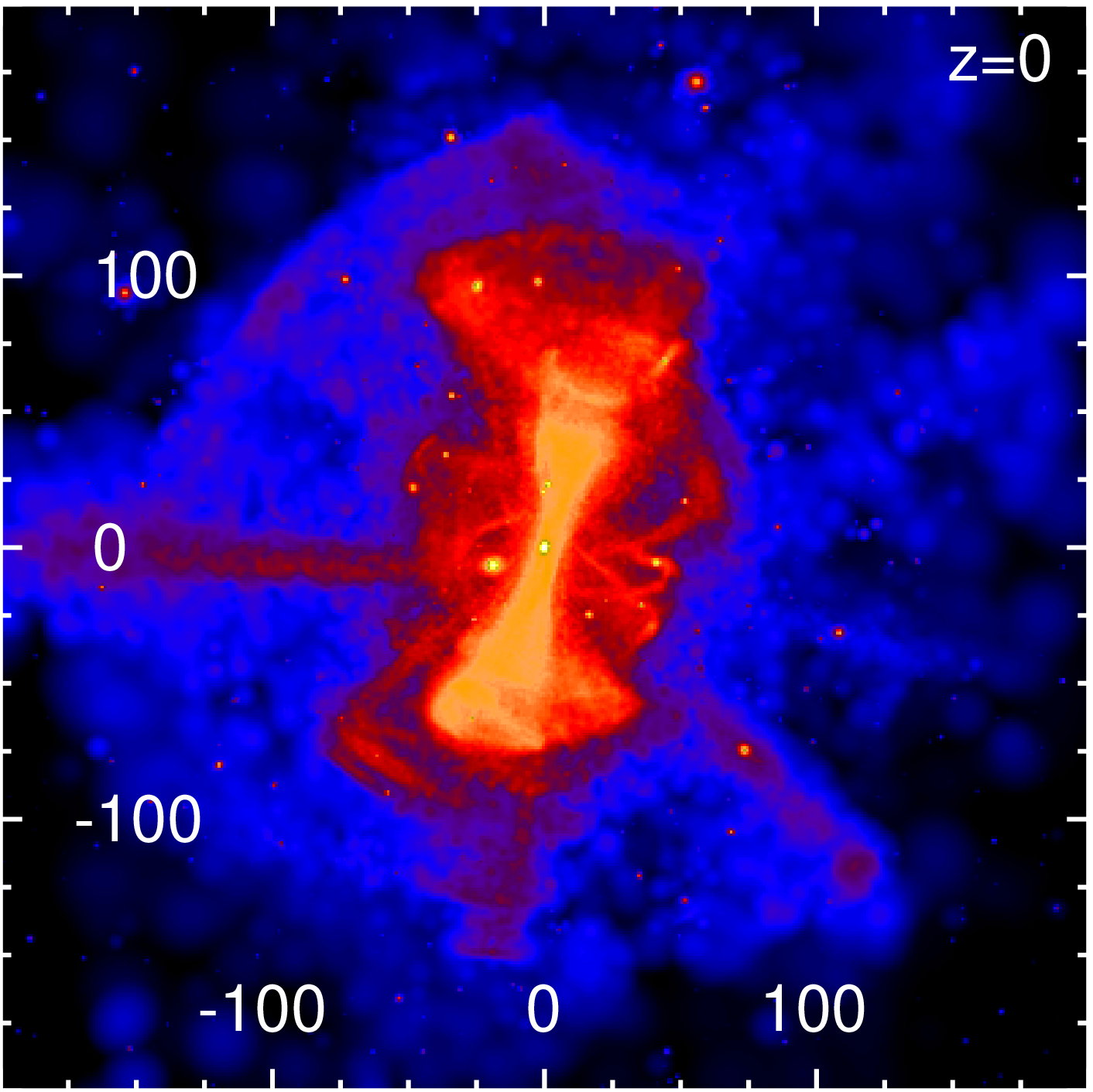} &
\includegraphics[width=59mm]{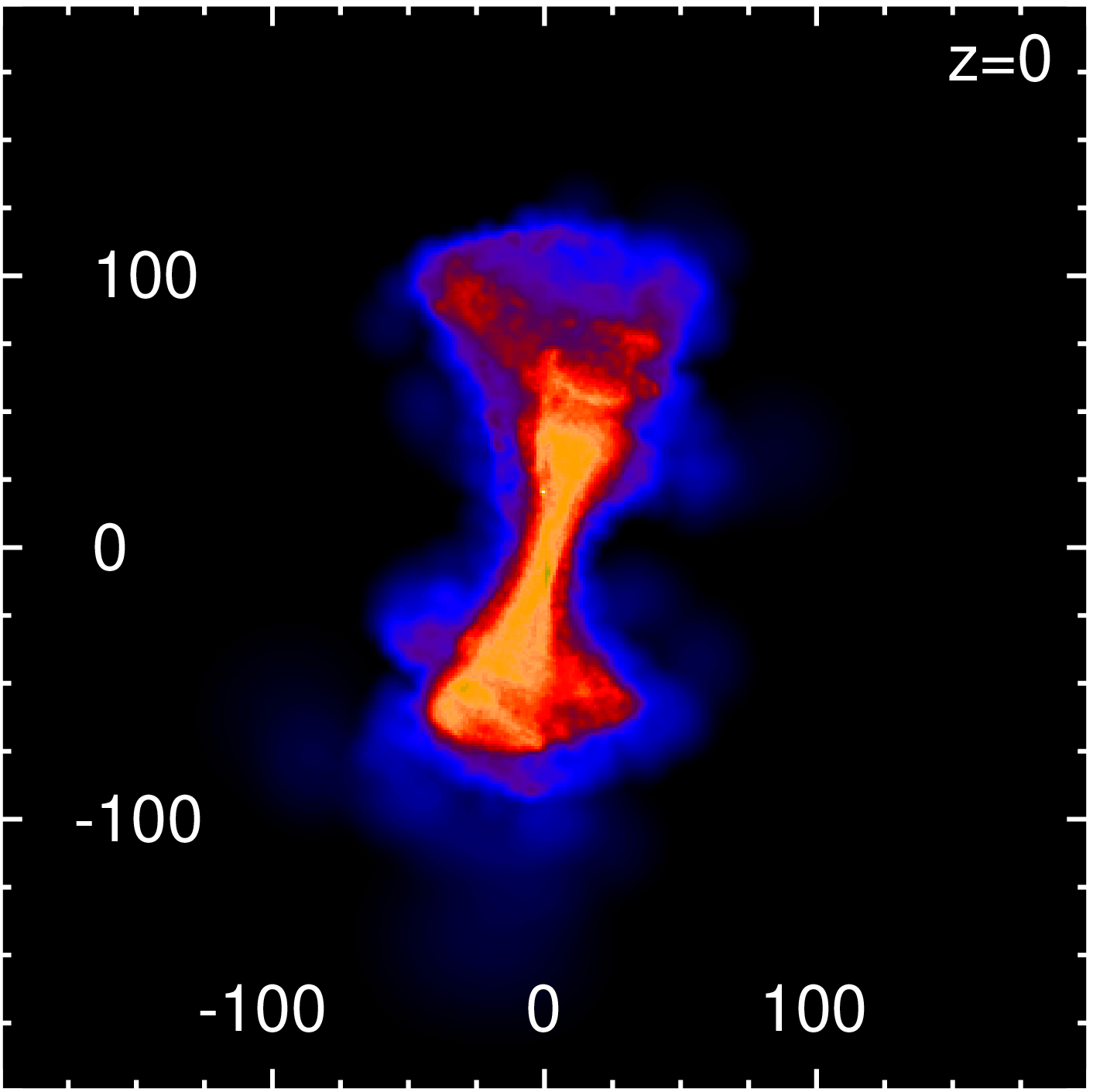} \\

\end{array}$
\end{center}
\caption{Projected mass surface density of the VLII dark matter halo ({\it left panels}) and its stellar halo ({\it central panels}), at redshift 2 ({\it top}), 1 ({\it middle}) and today ({\it bottom}). The image size is 400 physical kpc in all panels. For comparison, the $r_{200}$ radius of the VLII dark matter halo is 94 kpc, 180 kpc, and 400 kpc at these 3 epochs, respectively. The color scale (from dark to bright) ranges from  $2 \times 10^5\,\smsun$ to $10^9\,\smsun$ for 
the dark matter, and from $10^{-2}\,\smsun$ to $2 \times 10^8\,\smsun$ for the stars. {\it Right panels:} the disruption of the stellar 
component of an individual satellite tagged at infall ($z=2$).
}
\label{fig2}
\vspace{+0.3cm}
\end{figure*}

Figure~\ref{fig2} shows an image of the projected VLII dark matter and stellar halo mass density at 3 different redshifts. The disruption of the stellar component of an individual satellite is depicted in the right panels, where particles initially tagged in one subhalo at infall (the system reaches its $M_{\rm h}$ at $z=2$) are violently stripped by tides and form shell- and stream-like structures.  The bottom left and middle panels make it clear how the adopted steep stellar mass-halo mass relation offers an empirical solution to the ``missing satellite problem''. While the dark halo exhibits a vast amount of surviving self-bound substructure, only the most massive subhalos form enough stellar mass at infall to still be visible in the stellar halo today. 

\subsection{Surviving satellites}
\label{sec:satellites}

Surviving satellites in the $z=0$ VLII snapshot were identified by considering all particles that fall within the tidal radius of a bound subclump (identified in the dark matter density field by the 6DFOF group finder, see \citealt{die06}). We then performed a peculiar velocity cut of $V_p \leq 2 V_{\rm max}$ in order to remove particles belonging to the main halo that just happen to be flying through the given satellite at that time. The total luminosity of each satellite was computed by summing up the contributions of all stellar particles linked to it, and its metallicity determined from their mass-weighted average. Satellites of absolute magnitude $M_V$ located closer than the SDSS completeness limit $r_{\rm max}$ \citep{tol08} from the main halo center,
\begin{equation}
\label{eq:det}
r_{\rm max} = \left(\frac{3}{4\pi f_{\rm DR5}}\right)^{1/3} \times 10^{(-0.6M_V-5.23)/3}~{\rm Mpc},
\end{equation}
were flagged as observable in current surveys ($f_{\rm DR5}$ = 0.194 here is the sky covering fraction of the SDSS DR5).  Only 65 of the 1,925 surviving satellites present today in the VLII stellar halo are bright enough to be detected by an all-sky SDSS-like survey, out to a galactocentric distance of 280 kpc, corresponding to the Milky Way virial radius \citep{xue08}. 
Note that, as the SDSS detection efficiency is nearly independent of central surface brightness for $\mu_{0,V}\lta 30$ mag arcsec$^{-2}$ \citep{kop08}, 
no surface brightness threshold has been applied to our sample. We have checked that all but 13 of the observable VLII satellites have a central surface brightness
of at least 30 mag arcsec$^{-2}$. The 13 low surface brightness systems are all faint ($M_V\gta -5$) and have half-light radii $r_{\rm eff}$ close to the 
gravitational softening scale of the simulation, i.e. they have been artificially puffed up. We have estimated that, without such a numerical effect,        
these subhalos would actually lie above the SDSS surface brightness limit. Our prescriptions do not therefore produce a large population of low surface brightness ``stealth'' 
satellites awaiting discovery in the MW halo, as suggested by \citet{bul10}.

The stellar masses today of the observable VLII satellites range from $2,000\,\msun$ to $10^8\,\msun$, and their dark masses from $4\times 10^6\,\msun$ to $3\times 10^9\,\msun$. They have mass-to-light ratios that range from about $20$ to $10^5$, and infall redshifts mostly between 1.5 and 8 (with a few falling in later, between redshifts 0.6 and 1.1). Their mass-to-light ratios have decreased by a factor ranging between a few and 20, as their extended dark matter halos have been tidally stripped more than their compact luminous components. The ten most luminous satellites ($L>10^6\,\lsun$) in the simulation are hosted by subhalos with peak circular velocities today in the range $V_{\rm max}=10-40\,\kms$ that have shed between 80\% and 99\% of their dark mass after being accreted at redshifts $1.7<z<4.6$. Their $V_{\rm max}$ values have decreased by a factor of 2.5 on the average (and up to 6 in individual cases) after accretion, so their present-day $V_{\rm max}$ is not a good indicator of the subhalo maximum mass.
 
\begin{figure*}
\centering
\includegraphics[width = 0.49\textwidth]{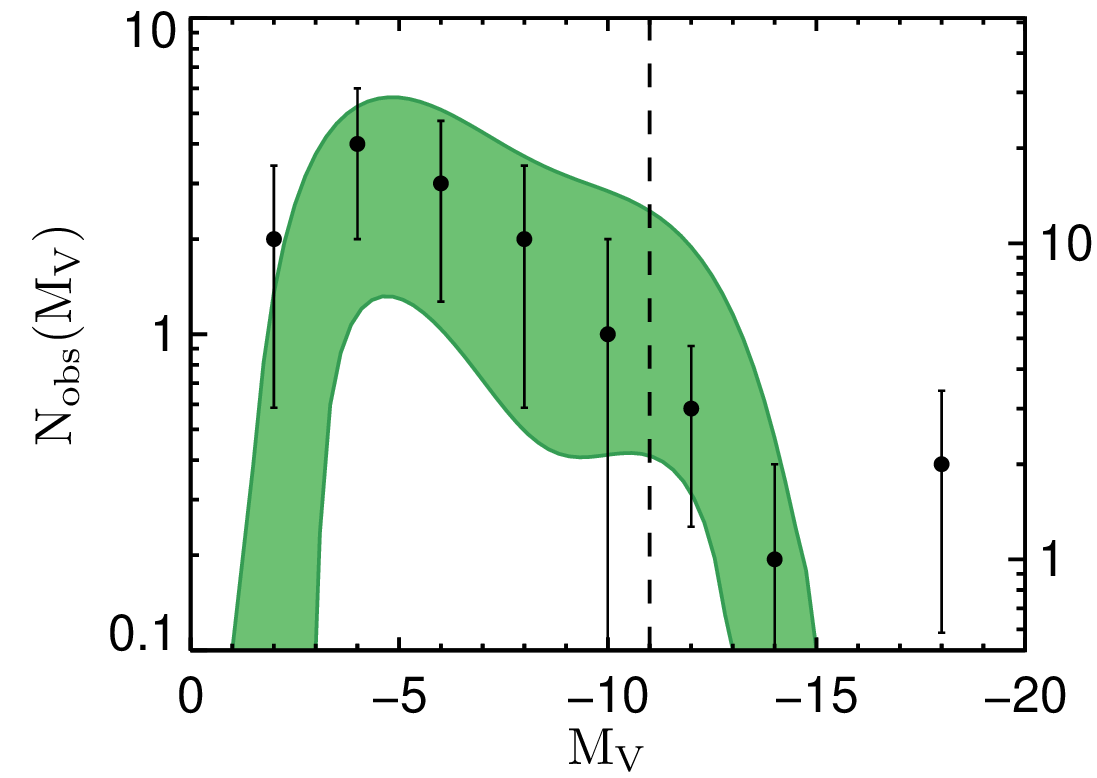}
\includegraphics[width = 0.49\textwidth]{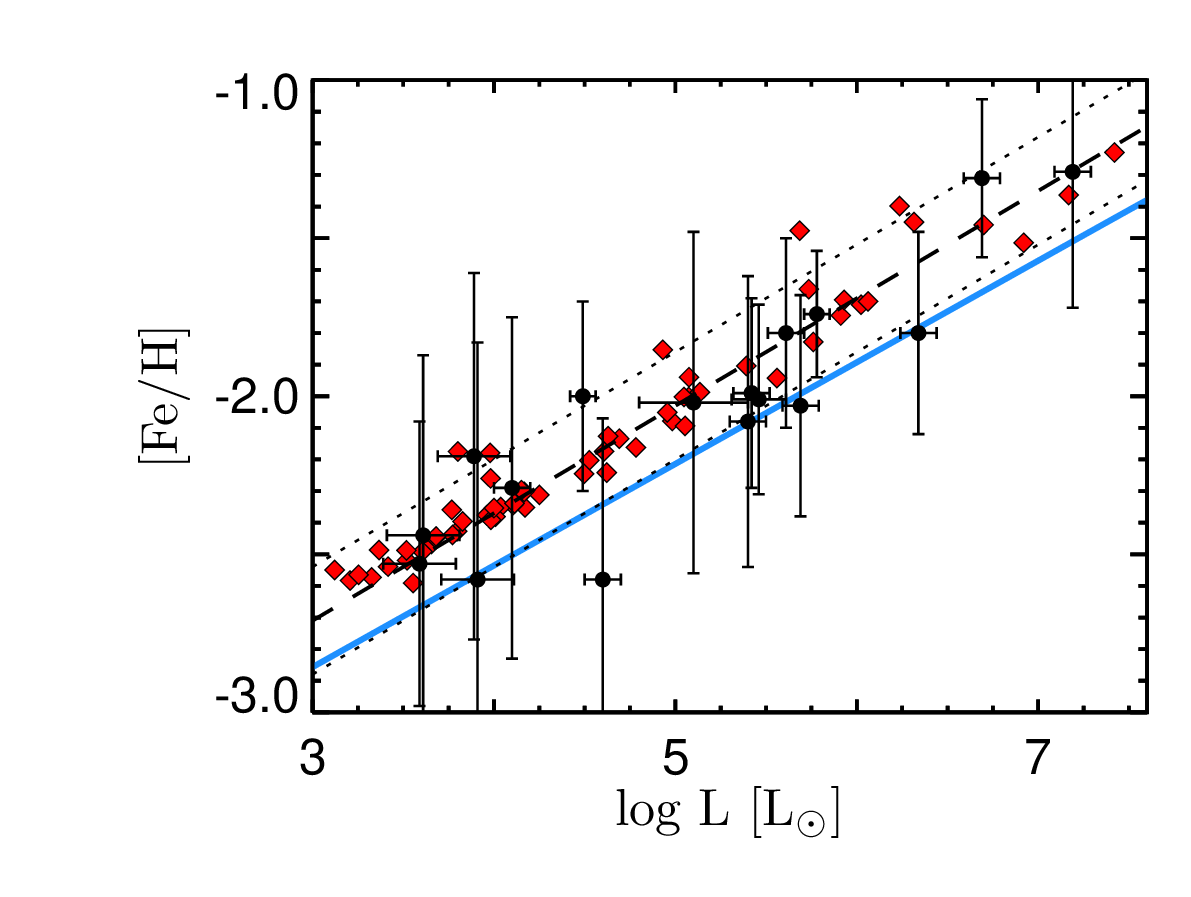}
\caption{
{\it Left panel:} Luminosity function of surviving satellites out to 280 kpc from the host halo center ({\it shaded in green}). Only galaxies that satisfy the magnitude-distance visibility SDSS criterion (see main text) are included. The shaded region for satellites fainter than $M_{\rm V} = -11$ spans the difference in number of dSphs falling within the SDSS DR5 footprint from 2,000 randomly oriented sky projections of the VLII halo {\it Black dots:} differential luminosity function of known MW satellites \citep[see][]{kop09}. Error bars are indicative of Poisson number statistics in each magnitude bin. The lack of Magellanic Cloud equivalents in the VLII halo is apparent in the bright-end discrepancy between model and data. Note the different scales for the number of satellites fainter (left y-axis) and brighter (right y-axis) than $M_{\rm V} = -11$, introduced to produce a continuous luminosity function despite the SDSS sky coverage of $f_{\rm DR5} = 0.194$.
{\it Right panel:} Evolution of the adopted metallicity--luminosity relation after tidal stripping and dimming of the stellar populations with time. The original prescription (eq. \ref{eq:met.lum}) assuming a mass-to-light ratio of unity is shown by a simple power-law ({\it solid blue line}). The stripping of stars from infalling satellites and the aging of stellar populations shifts luminosity to the left at a fixed value of [Fe/H]. {\it Red diamonds:} VLII satellites. The dashed and dotted lines show the observed metallicity--luminosity relation and 1$\sigma$ spread from \citet{kir08}. {\it Black circles}: Milky Way dSphs; data from \citet{kir08} and references therein.
}
\label{fig3}
\vspace{+0.3cm}
\end{figure*}

The differential satellite luminosity function is shown in the left panel of Figure~\ref{fig3}, plotted against the luminosity function of observed MW dwarfs \citep[see also works by][]{kop09, mac10}. As in \citet{kop09}, we use different scales for the number of satellites brighter and fainter than $M_{\rm V} = -11$, introduced to produce a continuous luminosity function despite the limited SDSS DR5 sky coverage of $f_{\rm DR5} = 0.194$. The shaded region to the left of M$_{\rm V} = -11$ spans the 1$\sigma$ extent in the number of faint, but detectable satellites that fall within the SDSS DR5 footprint in each of 2,000 randomly oriented sky projections of the VLII stellar halo. In the case that the satellites are isotropically distributed, our exercise should produce a result equivalent to a scaling of the number counts of observed faint satellites by the sky coverage fraction of the SDSS survey. Satellites to the right of the dashed line, corresponding to $M_{\rm V} = -11$ are always detectable. The shaded region to the right of M$_{\rm V} = -11$ covers the 1$\sigma$ Poisson errors in each bin there.
The lack of Magellanic Cloud equivalents in VLII is responsible for the discrepant bright end of the luminosity function, as the simulation does not contain satellites more luminous than $M_V \lesssim -14$. Studies suggest that satellites this bright are rare around Milky Way-like galaxies \citep{boy10,liu10}.
In the right panel of the same figure, we plot the luminosity--metallicity relation of surviving satellites compared to data for MW dSphs \citep[][and references therein]{kir08} and their best-fit model to the data. The relation assumed at infall (eq. ~\ref{eq:met.lum}) has been modified by tidal mass losses and by the aging of stellar populations.

\begin{figure*}
\centering
\includegraphics[width = 0.49\textwidth]{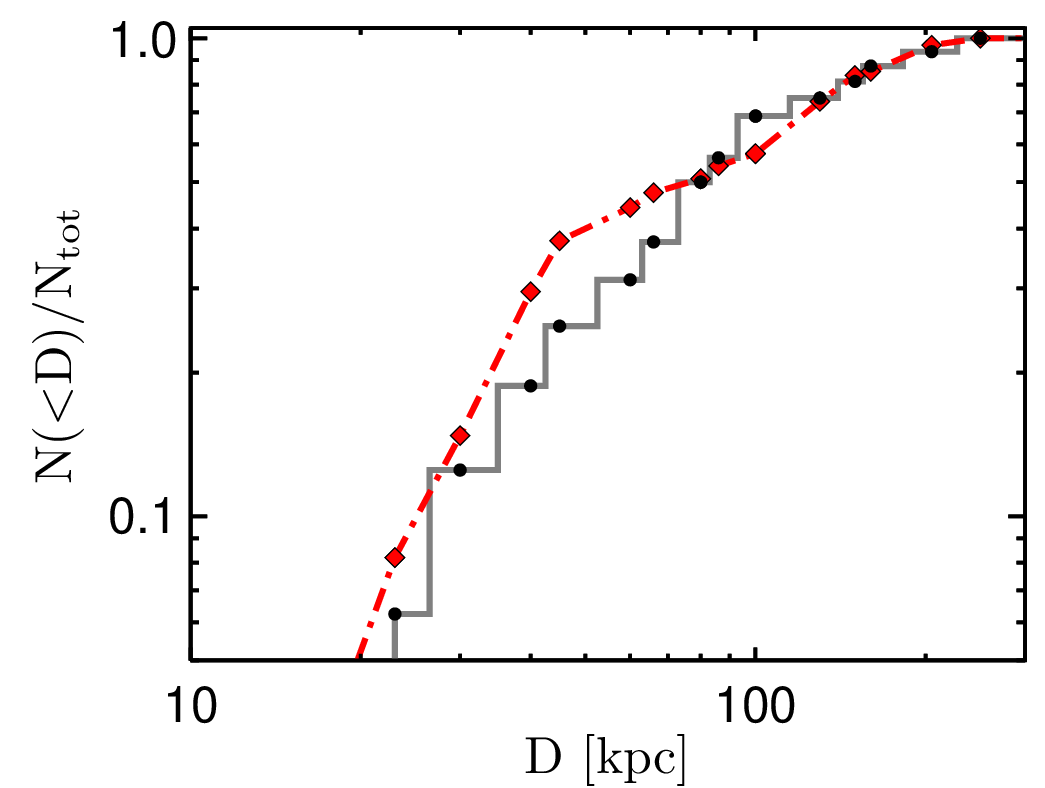}
\includegraphics[width = 0.49\textwidth]{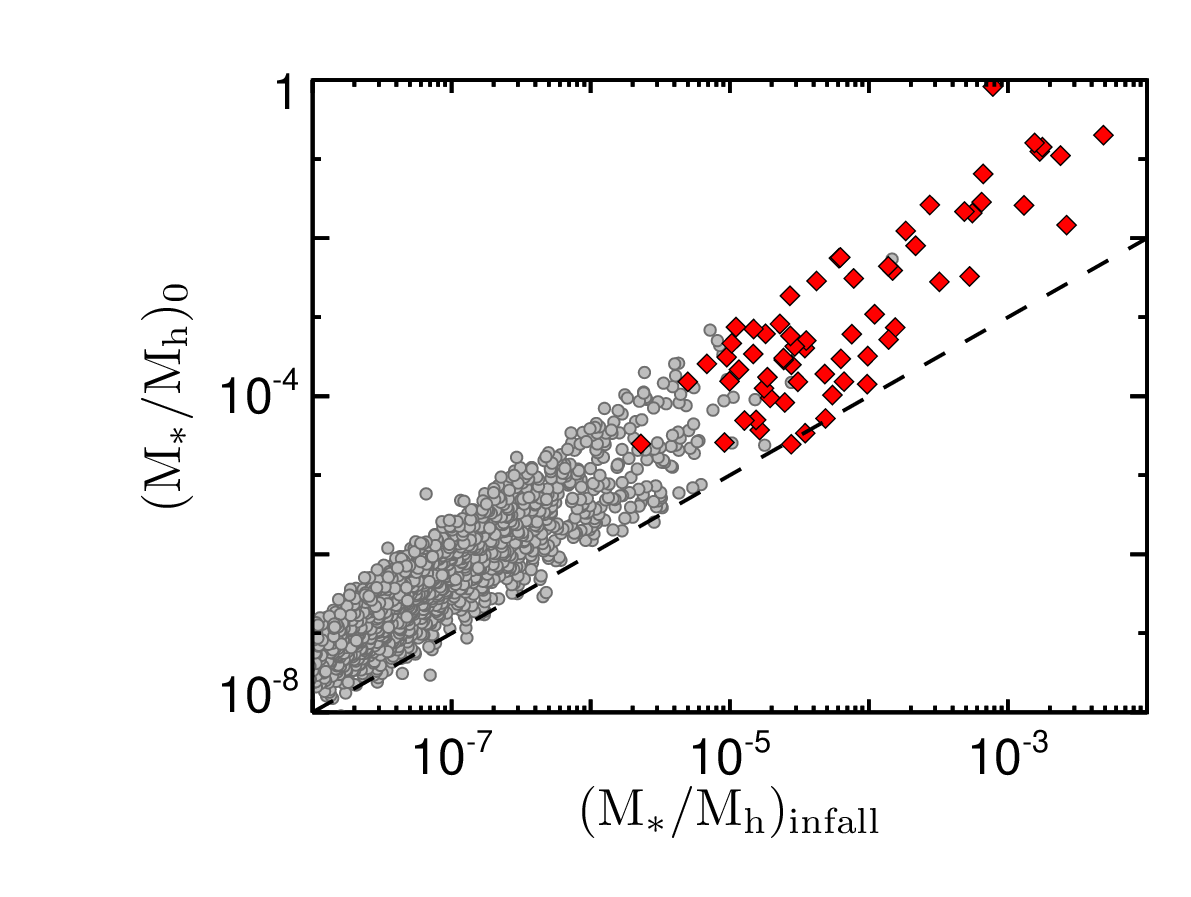}
\includegraphics[width = 0.49\textwidth]{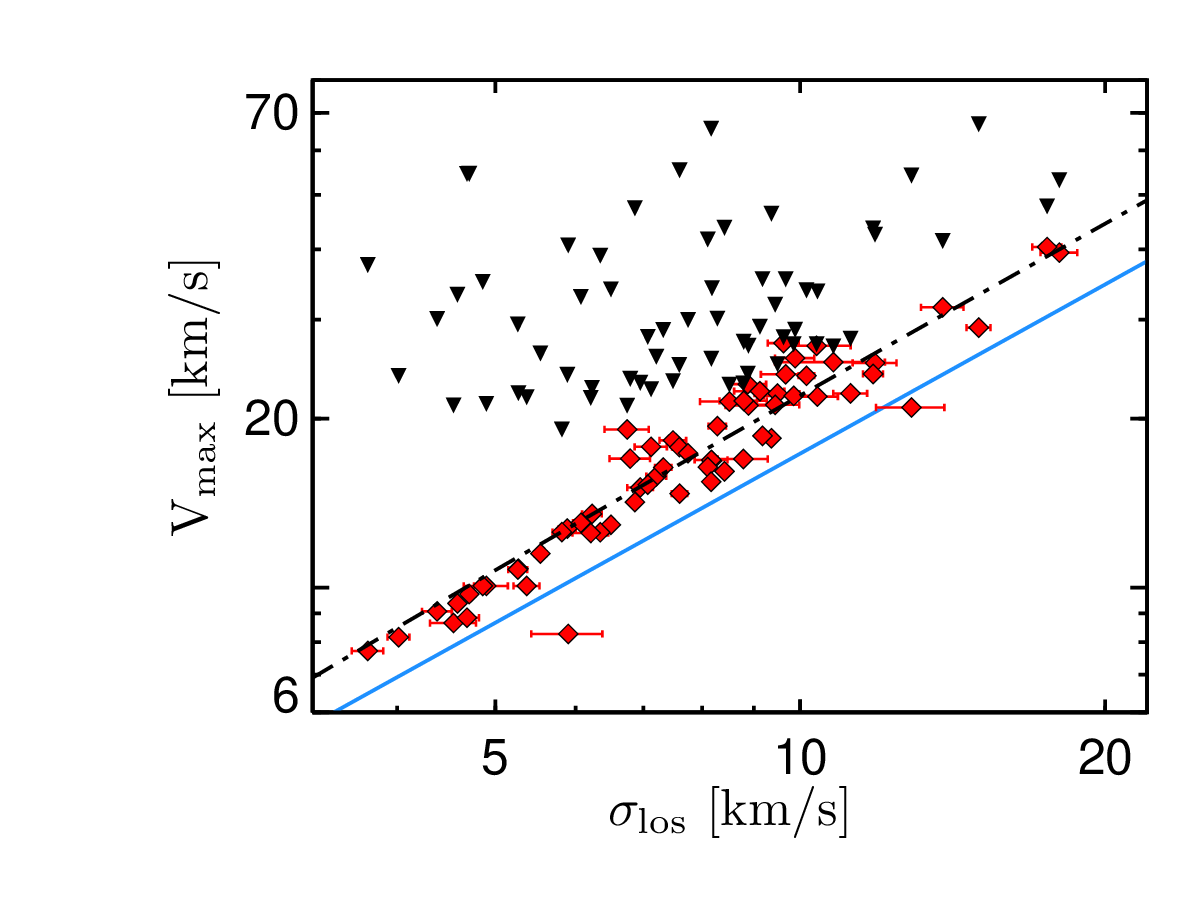}
\includegraphics[width = 0.49\textwidth]{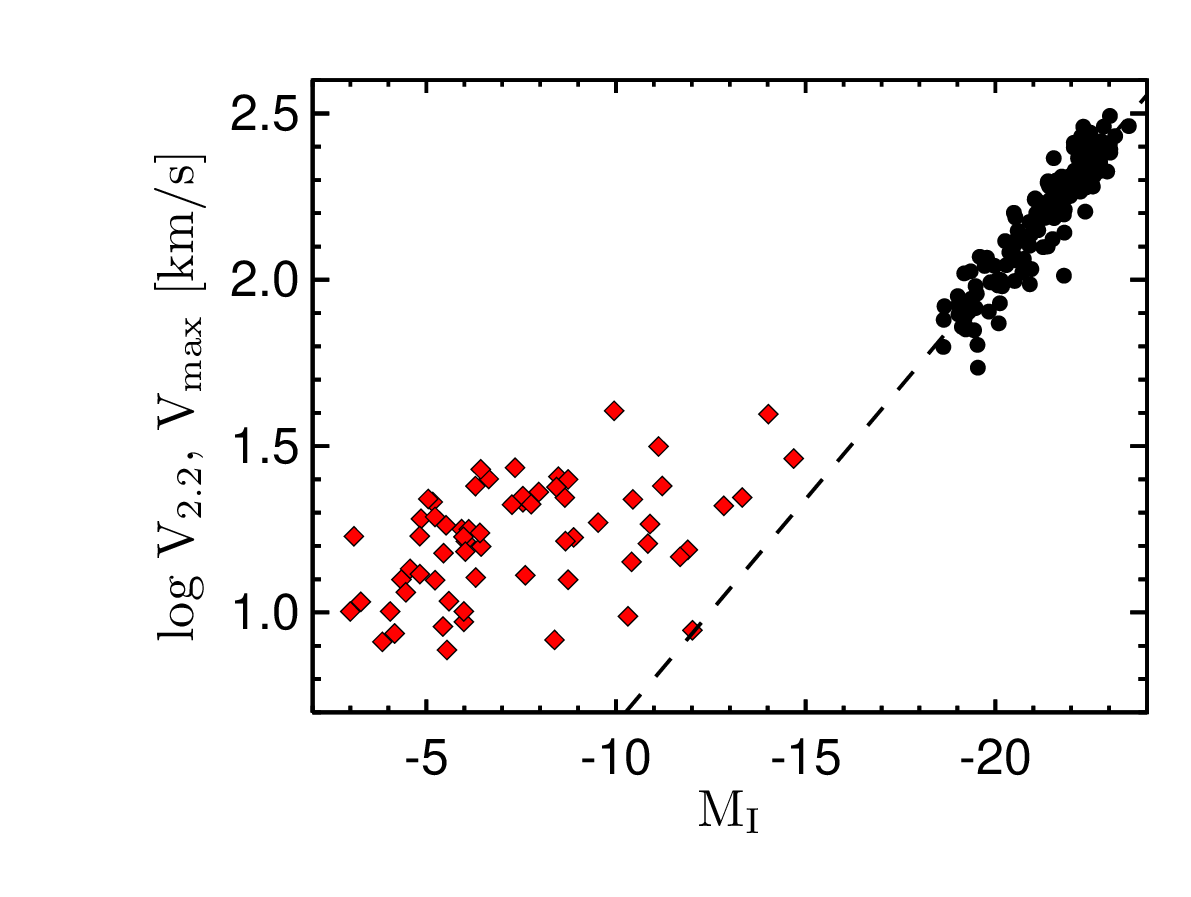}
\caption{{\it Left top panel:} Cumulative distribution of galactocentric distance of VLII satellites fainter than $M_V \ge -12$, corresponding to the dSph satellites discovered by the SDSS. {\it Black circles:} Milky Way dSphs. {\it Red diamonds:} VLII satellites.
{\it Right top panel:} Evolution of the stellar mass-halo mass relation from the infall epoch to the present in VLII surviving subhalos. {\it Red diamonds:} satellites detectable in an all-sky SDSS-like survey. {\it Grey dots:} all remaining (currently undetectable) satellites. The apparent increase in the
star formation efficiency is caused by the preferential tidal stripping of the more diffuse dark matter component relative to the inner, more tightly bound stellar particles.
{\it Left bottom panel:} Correspondence between 1-D line-of-sight velocity dispersion $\sigma_{\rm los}$ and 3-D maximum circular velocity $V_{\rm max}$ for observable VLII dSphs. {\it Solid blue line:} $V_{\rm max} \propto \sqrt{3}\sigma_{\rm los}$. {\it Red diamonds:} VLII detectable satellites. Error bars indicate 1$\sigma$ variation over the 11 measurements of each profile used in our kinematic study. {\it Dashed line:} $V_{\rm max} \propto 2.2\sigma_{\rm los}$. {\it Black triangles:} Highest value of $V_{\rm max}$ ever achieved by each satellite.
{\it Right bottom panel:} The $V_{\rm max}-$I band magnitude relation for surviving satellites ({\it red diamonds}) compared to to the Tully-Fisher relation of a broadly selected galaxy sample ({\it black dots}) from the SDSS \citep{piz07}. The dashed line gives the extrapolation of the best fit to the galaxy data. VLII dSphs clearly do not follow the relation.
}
\label{fig4}
\vspace{+0.0cm}
\end{figure*}

\begin{figure}
\centering
\includegraphics[width = 0.49\textwidth]{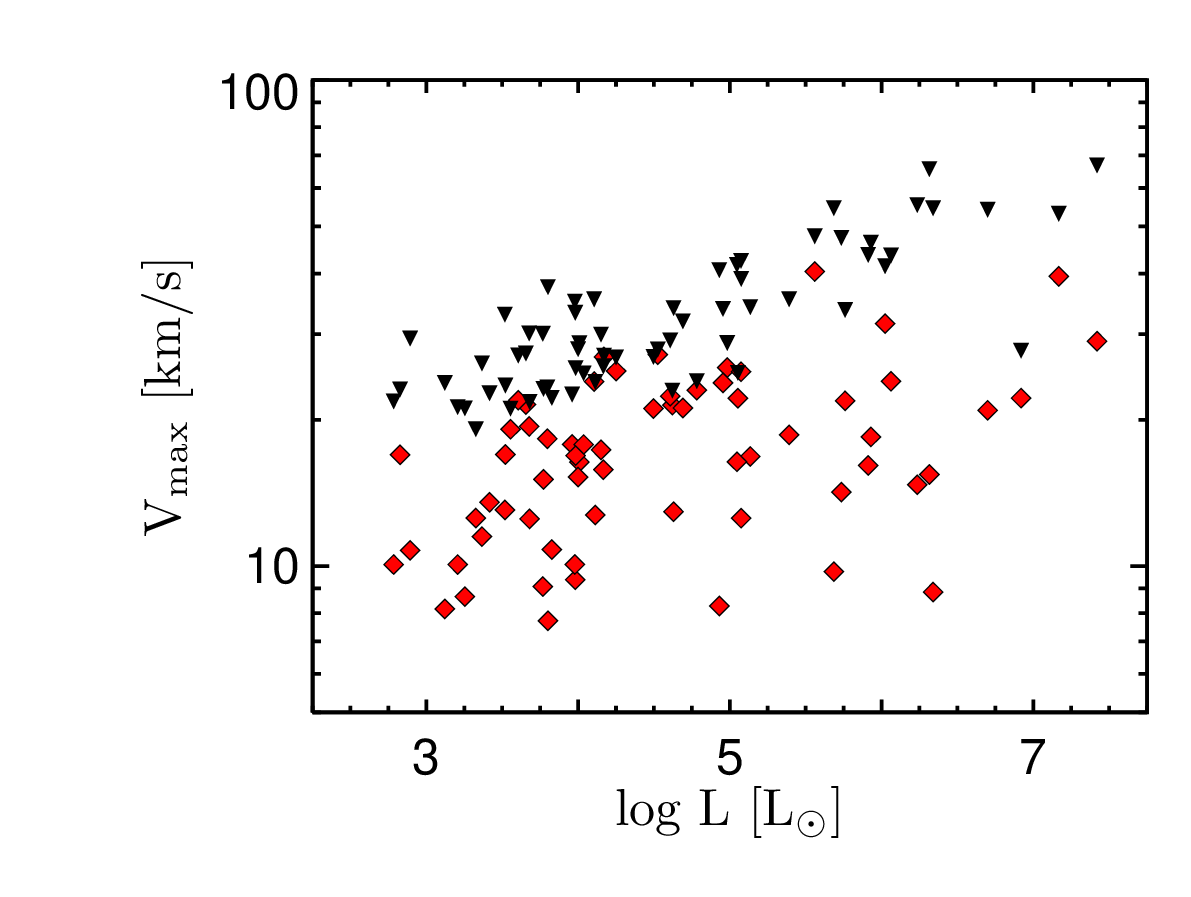}
\caption{
Comparison between the values of the 3-D maximum circular velocity $V_{\rm max}$ at infall and today as a function of the satellite's present-day luminosity, for observable VLII dSphs. {\it Red diamonds:} 
values of $V_{\rm max}$ today. {\it Black triangles:} the highest values of $V_{\rm max}$ ever achieved by each satellite. For the 10 most luminous satellites today, the $V_{\rm max}$ values have decreased 
by a factor of 2.5 on average (and up to 6 in individual cases) since accretion.
}
\label{fig5}
\vspace{+0.0cm}
\end{figure}

In Figure~\ref{fig4}, the cumulative galactocentric distance distribution of VLII luminous subhalos is compared to that of MW satellites fainter than $M_V \ge -12$. Even though the final spatial distribution of dwarfs depends on their accretion and orbital history, the two distributions track each other reasonably well, though VLII shows an excess of close-in satellites compared to the observations. 
The figure also shows (right top panel) the present-day stellar mass--subhalo mass relation. Much like the metallicity--luminosity relation in Figure~\ref{fig3}, it has been modified from its infall value by tidal mass losses. As accreted systems lose a disproportionate amount of their dark mass, the apparent (inferred today) star formation efficiency of dSphs may exceed the true one by as much as a factor of 100.
We also explore (bottom left panel of Fig. \ref{fig4}) how the average 1-D line-of-sight velocity dispersion $\sigma_{\rm los}$ of dSphs (an observable quantity) compares to the peak (3-D) circular velocity $V_{\rm max}$ of the system's host halo (a quantity of theoretical interest). Early attempts at this conversion assumed a proportionality factor of $\sqrt{3}$ motivated by the isotropy of these systems \citep{moo99}. Our model suggests that the baryonic tracers are deeply embedded in their dark matter potentials today, resulting in higher values of $V_{\rm max}$, proportional to 2.2$\sigma_{\rm los}$.\footnote{This $\sim$30\% difference in $V_{\rm max}$ corresponds to a factor of 2 increase in total mass, $M \propto V_{\rm max}^3$.} The difference is even more pronounced when comparing to the highest $V_{\rm max}$ values these systems ever achieved. 
Shown in the Figure as downward pointing triangles, these values indicate the amount of tidal mass loss that the subhalos hosting dSphs have undergone between infall and today. In Figure \ref{fig5}, we re-iterate this point by plotting the time evolution of $V_{\rm max}$ as a function of satellites' present-day luminosities.

Finally, we show how VLII dSphs compare to the Tully-Fisher relation \citep[data from][]{piz07} for disk galaxies (Fig. \ref{fig4}, bottom right panel). Just like the MW dSph population, our star formation prescription fails to produce dwarfs that lie on the extrapolated Tully-Fisher relation. The flattening of the distribution of spheroidal galaxies to a constant velocity value in comparison to field spiral galaxies is reminiscent of the ``common mass scale'' seen in the M$_{\rm 300}$ mass function (see Fig.~\ref{fig9}), in which all currently detectable dSphs cluster around the value M$_{\rm 300} \sim 10^7\, \msun$. As we shall see below, in 
our model the ``common mass scale'' is the result of a luminosity bias, and future detection of fainter dSph satellites would break it and reveal 
lighter dark matter subhalos hosting stellar systems.

We now turn to the structural properties of the surviving VLII dwarfs. For all satellites observable in an all-sky SDSS-like survey, we calculate the circularly-averaged surface brightness seen by a galactocentric observer and fit it with a 2-D projected King profile \citep{kin62},
\begin{equation}
I_*(R)=k\left[\left(1+\frac{R^2}{r_k^2}\right)^{-1/2}-\left(1+\frac{r_t^2}{r_k^2}\right)^{-1/2}\right]^2,
\end{equation}
where $I_*(R)$ is the surface brightness at radius R from the satellite center, k is the profile normalization, $r_k$ is the core radius of the satellite, and $r_t$ is its tidal (cutoff) radius. This formula provides a good fit to the surface brightness profiles of MW dSphs \citep{mat98}, 
and has the advantage of having an analytical de-projected 3-D form in spherical geometry. Our use of the King profile as a fit to the projected surface brightness of VLII satellites instead of, say an exponential profile, is motivated by its standard use in published studies of dSph mass estimation. We find that the luminous component of the observable simulated dSph is well fit by a King profile across the full range of observable luminosities. Note that, while the tidal radii in the luminous component do not correspond to the physical truncation scale in the satellites' dark matter extent, the two scales are indeed comparable. On average, the ratio between the tidal radius in the luminous component and the tidal radius in the dark matter extent varies between 0.5 and 1 for satellites more luminous than $10^6 L_{\odot}$, while it is between 0.3 and 0.8 for fainter systems.

\begin{figure*}
\centering
\includegraphics[width = 0.49\textwidth]{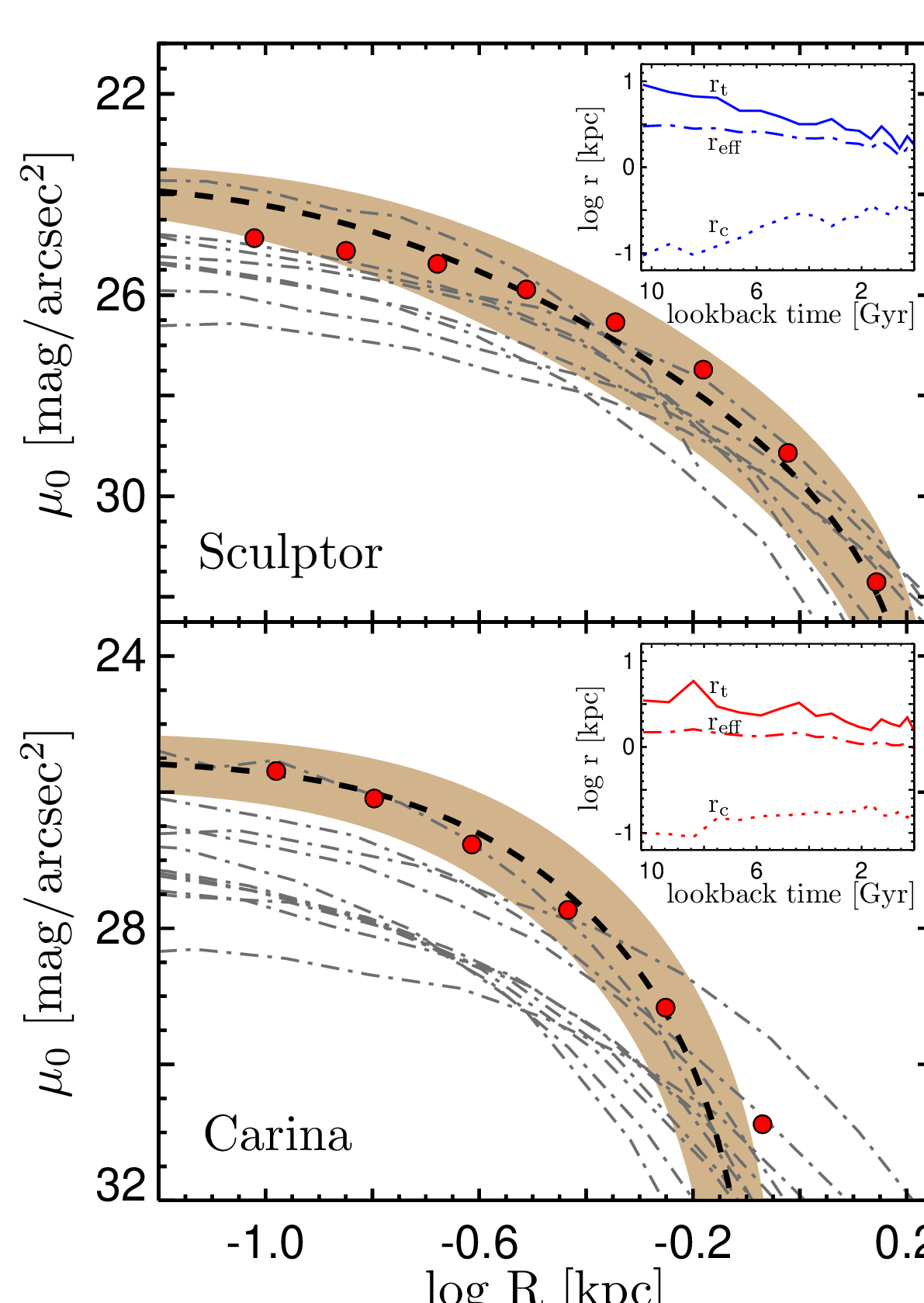}
\includegraphics[width = 0.49\textwidth]{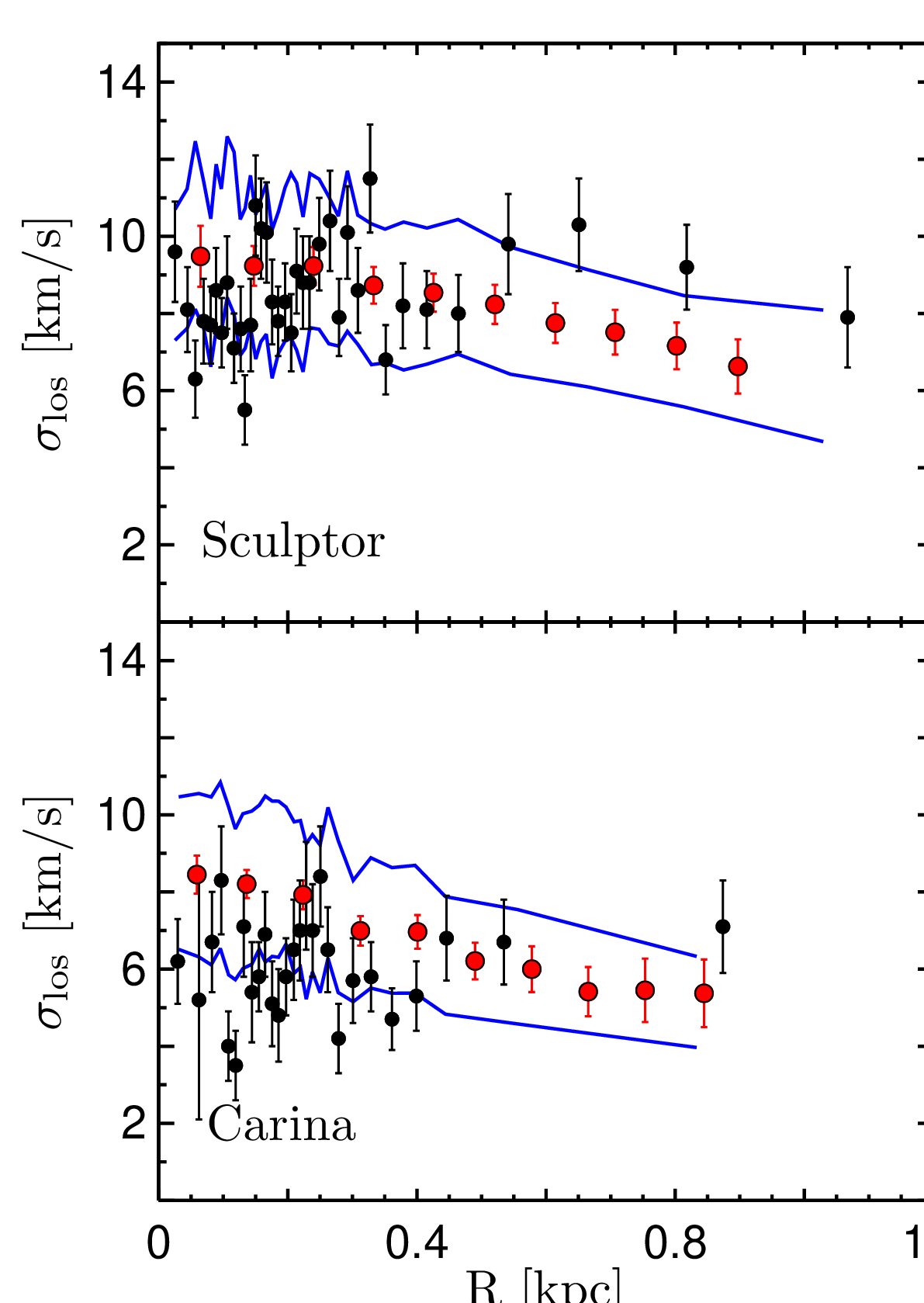}
\caption{Surface brightness and velocity dispersion profiles of two VLII  satellites ({\it red dots}) closely resembling the Milky Way dSph Sculptor and Carina. Error bars are indicative 
of 1$\sigma$ Bootstrap re-sampling errors in each bin. Simulated surface brightness and kinematics should be compared only for $R>80$ pc, corresponding to twice the VLII softening length. The observed surface brightness profiles are plotted with dashed lines and the uncertainty is shown by the shaded region. Brightness fits for Sculptor and Carina are taken from \citet{mat98}, while the kinematic data ({\it black dots with error bars}) are from \citet{wal09}. Also shown is the diversity in brightness profiles among VLII satellites with luminosities within $\pm$80\% of Sculptor's and Carina's ({\it grey dot-dashed lines}). The insets follow the evolution of the King profile structural parameters for the fiducial satellites from the time of `tagging' (infall) until today. {\it Solid blue lines:} the 95\% confidence region in the kinematic profiles 
where the data are sparsely sampled as the observations.
}
\label{fig6}
\vspace{+0.3cm}
\end{figure*}

A comparison between the surface brightness and line-of-sight velocity dispersion profiles of two dSph satellites (Sculptor and Carina) and individual VLII counterparts is shown in Figure~\ref{fig6}. Both surface brightness and kinematics should be compared only at radii larger than twice the VLII softening length of 40 pc, as the artificial ``heating"  of particles inside this region from numerical resolution effects would give a poor match to the data. In the surface brightness plot, we show the diverse light profiles of all VLII dwarfs that have a luminosity within $\pm 80\%$ of the corresponding MW satellite. Equally diverse is the span of velocity dispersion profiles these dwarfs have, with values ranging between 4$\kms$ and 15$\kms$, pointing to the differences between the potential wells in which these systems reside. The insets in the figure show the evolution of the luminous systems' parameters (King profile core and tidal radii and half-light radii) from `tagging' (infall) until today. The behavior seen here is representative of the other SDSS detectable dwarfs in VLII - the core radii increase with time (though they remain close to the simulation softening length), while the half-light and tidal radii decrease as the satellites lose stars from their outskirts to the main halo. In the velocity dispersion plots, we try to reproduce the `noisy' observed profile by sparsely sampling the simulation data. Using the same number of stellar particles in each corresponding bin as the one in the observed profile (M. Walker, private communication), we note that the scatter in the simulated profile increases significantly.

\begin{figure*}
\centering
\includegraphics[width = 176mm]{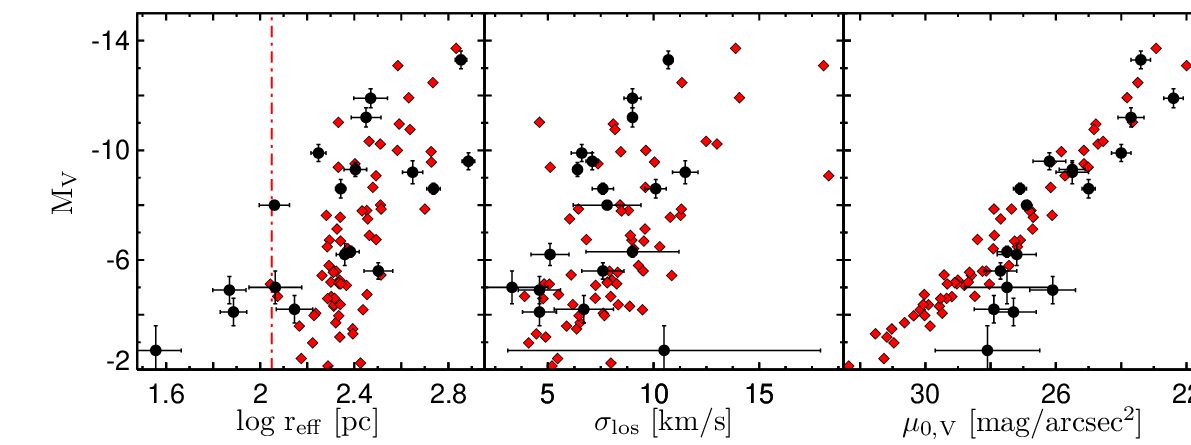}
\caption{Structural properties of surviving VLII satellites. Plotted against V-band absolute magnitude are their half-light radii ({\it left panel}), line-of-sight velocity dispersions ({\it middle panel}), and central surface brightnesses ({\it right panel}). {\it Red diamonds:} observable VLII 
dwarfs. {\it Black circles with error bars:} Milky Way dSph. Data compiled as follows: {\it $M_V$}: {\it classical dwarfs:} \citep{mat98}, except for {\it Leo II} \citep{col07}, {\it ultra-faint dwarfs:} \citep{mar08}, except for {\it Hercules} \citep{san09} and {\it Leo T} \citep{dej08}; {\it $r_{\rm eff}$} and {\it $\sigma_{\rm los}$}: \citet{wol10}; {\it $\mu_{\rm 0, V}$}: {\it classical dwarfs;} \citep{mat98}, except for {\it Draco} \citep{mar08}, {\it ultra-faint dwarfs:} \citep{mar08}, except for {\it Leo T} \citep{irw07}. The dash-dot red vertical line in the left panel marks 2.8$\times$ the VLII softening length of 40 pc \citep[see][]{coo10}.}
\label{fig7}
\vspace{+0.3cm}
\end{figure*}

In a systematic analysis, \citet{str10} have shown that at least one simultaneously structural and kinematic counterpart to each of the five classical dSph with good kinematic and photometric data can be found in the Aquarius simulations. Our ability to obtain simultaneous brightness and kinematic fits for dSphs like Sculptor and Carina can be seen as a confirmation of their results and a success of our model, and is a prerequisite for testing the ``common mass scale'' of MW satellites. Structural parameters of the VLII satellite population (half-light radii and velocity dispersions measured directly from the particle data, and central surface brightnesses from the King model fits as a function of absolute magnitude) are compared to the observed properties of MW dSphs in Figure~\ref{fig7}. The agreement is reasonably good. Reproducing the properties of the luminous component of dSph galaxies is a necessary condition for generating realistic ``mock'' data to test the mass measurement technique. 

\section{Kinematics of dwarf spheroidal satellites}
\label{sec:kinematics}

\subsection{The spherical Jeans equation}

The Jeans equation is the first moment over velocity of the collisionless Boltzmann equation. It provides a link between the underlying potential of a pressure-supported system and the distribution function of a tracer population in equilibrium with it. In spherical geometry its radial component is:
\begin{equation}
r\frac{d\left(\nu_*\sigma_r^2\right)}{dr} = -\nu_*V_c^2-2\beta \nu_*\sigma_r^2,
\label{eq:jeans}
\end{equation}
where $\nu_*(r)$ is the luminosity density profile of the tracer stars, $\sigma_r$ is their radial velocity dispersion, and $\beta = 1 - \sigma_t^2/\sigma_r^2$ is the velocity anisotropy, i.e. the difference between the radial, $\sigma_r$, and the tangential, $\sigma_t^2 = (\sigma_\theta^2 + \sigma_\phi^2)/2$, components of the velocity dispersion. Here, $V_c(r)$ is the circular velocity profile of the system, a measure of the underlying gravitational potential $\Phi$, 
\begin{equation}
V_c(r)^2 = G\frac{M(<r)}{r} = -r\frac{d\Phi}{dr}.
\end{equation}

Since 3-D kinematic data for MW dSph stars are not available, one has to use a 1-D projection of the spherical Jeans equation along the line-of-sight between the observer and the studied system. This introduces an additional geometric effect involving the velocity anisotropy \citep{bin08}:
\begin{equation}
\sigma_{\rm los}^2(R) = \frac{2}{I_*(R)}\int_R^{\infty}\left(1-\beta\frac{R^2}{r^2}\right)\frac{\nu_*\sigma_r^2 r}{\sqrt{r^2 - R^2}}dr,
\label{eq:jeansproj}
\end{equation}
where $I_*(R)$ is the projected surface brightness profile of the tracer population.
Of all quantities involved in the projected Jeans equation, $\sigma_{\rm los}$ and $I_*(R)$ are readily determined observationally, and $\nu_*(r)$ can be obtained through a de-projection of the surface brightness profile, assuming it follows some analytical form - either a spherical \citet{kin62} or a \citet{plu11} profile. Both of these profiles adequately fit the light distribution of MW satellites and the choice of one or the other does not affect the results of our kinematic study \citep{str08}. The remaining parameters, namely $\beta(r)$ and $V_c(r)$, are not constrained observationally and have to be modeled appropriately.

We adopt a generalized NFW profile \citep{nav96} for the mass distribution of subhalos:
\begin{equation}
\rho(r) = \frac{\rho_s}{(r/r_s)^{\gamma}(1+r/r_s)^{\delta}}, 
\end{equation}
where $\rho_s$ gives the density normalization, $r_s$ is the scale radius, and $\gamma$ and $\delta$ define the inner and outer slopes of the density fall-off. The fiducial NFW form has $\gamma = 1$ and $\delta = 2$. To allow for a range of inner density cusps as well as outer tidally-modified profiles, we allow these two parameters to vary in the intervals $0.7 \lesssim \gamma \lesssim 1.2$ and $2 \lesssim \delta \lesssim 3$ \citep{str07}. We have verified that the upper limit on the outer slope $\delta$ is high enough to accommodate the profiles of even the most massive VLII subhalos, whose density profiles fall off faster than those of field halos in the tidally-stripped outer regions \citep{die08}.

The velocity anisotropy parameter $\beta$ is largely unconstrained by current observations. According to its definition, $\beta$ can vary between $-\infty$ and 1, with $\beta = -\infty$ when $ \sigma_{\rm tot} = \sigma_t$ and $\beta = 1$ when $\sigma_{\rm tot} = \sigma_r$. Available data are often in the form of 1-D line-of-sight velocity dispersion, insufficient to constrain $\sigma_r$ and $\sigma_t$ individually. These two quantities are well determined only at the very center and very edge of a system, since $\sigma_{\rm los}(0) = \sigma_r$ and $\sigma_{\rm los}(r_t) = \sigma_t$. In between these two extremes, the mass of a dSph enclosed within a radius close to the system's half-light radius can be calculated such that the uncertainty from the velocity anisotropy term is minimized \citep{wol10}. At any other intermediate radius, the $\beta$ parameter will more strongly affect the mass measurement and must be estimated using a theoretical model. In principle, $\beta$ can be a function of radius within a particular system \citep{str07}. Previous studies have shown that marginalizing over a set of parameters that define a general anisotropy profile does not constrain the mass distribution better than assuming $\beta$ to be constant with radius \citep{wal09}. The latter case, however, greatly simplifies the solution to the spherical Jeans equation \citep{mam05}:
\begin{equation}
\nu_*\sigma_r^2(R) = GR^{-2\beta} \int_R^{\infty}r^{2\beta-2}\nu_*(r)M(r)dr.
\end{equation}

The standard procedure in mass modeling of pressure-supported systems is to generate a theoretical velocity dispersion profile that fits the observed one, and to marginalize the solution over all free parameters but the density profile normalization. While most model parameters are largely unconstrained by this fitting procedure, the density profile normalization (defined as the total amount of mass enclosed in a fixed aperture radius, typically 300 pc or 600 pc) is very well determined by the data \citep{str08, wal09}. It is this analysis that yields the so-called ``common mass scale'' of MW dSph satellites.

\subsection{Applications to VLII dwarf satellites}
\label{sub:vl2kin}

We apply the mass measurement technique above to the simulated data using the following procedure:
\begin{enumerate}
\item Obtain $\sigma_{\rm los}(R)$ and $I_*(R)$ for a satellite of interest;
\item Choose the density profile and anisotropy parameters $\rho_s$, $r_s$, $\gamma$, $\delta$, and $\beta$;
\item Integrate the Jeans equation (\ref{eq:jeans}) to solve for $\sigma_r(r)$;
\item Project $\sigma_r(r)$ along the line of sight (eq. \ref{eq:jeansproj});
\item Compare the solution to the measured $\sigma_{\rm los}(R)$;
\item Iterate (2 - 5) to find the model that best matches the kinematic data.
\end{enumerate}

Once the best fit mass density profile is determined, we integrate it to a given aperture radius, either 300 pc or 600 pc from the satellite center to obtain the total 
enclosed mass, M$_{\rm 300}$ or M$_{\rm 600}$, and compare it directly to the ``true" value in the simulation.
To obtain the ``mock'' observations of VLII dSphs in step 1, we place ``hypothetical observers'' in the simulation volume, and reduce computational time
by ``observing" each of the 65 detectable satellites only 11 times. One of the ``observers'' is always in the center of the VLII main host, while the other 
10 are randomly distributed at the same distance from the satellite as the galactocentric ``observer''. In this way we attempt to capture the anisotropy of subhalos' 
physical shapes and velocity ellipsoids.

Since integrating the Jeans equation requires a boundary condition on $\sigma_r(r_t)$, which is not known a priori from the data, the line-of-sight projected solution cannot be trusted across the full extent of the surface brightness fit. The observed line-of-sight velocity dispersion profile is therefore calculated in 10 linear distance bins out to 60\% of each satellite's tidal radius $r_{t}$. 
The simulated data used to determine each subhalo's kinematic profile do not have an associated observational uncertainty. To make a fair comparison with the derived mass estimates, ``observational'' errors should be consistent with the ones in the current literature \citep{wal09}. This is done by randomly assigning error values $\sigma_i$ ranging between 1 and 2 $\kms$ to the velocity dispersion data when evaluating the goodness of fit in equation (\ref{eq:like}) below.

\begin{figure*}
\centering
\includegraphics[width = 0.49\textwidth]{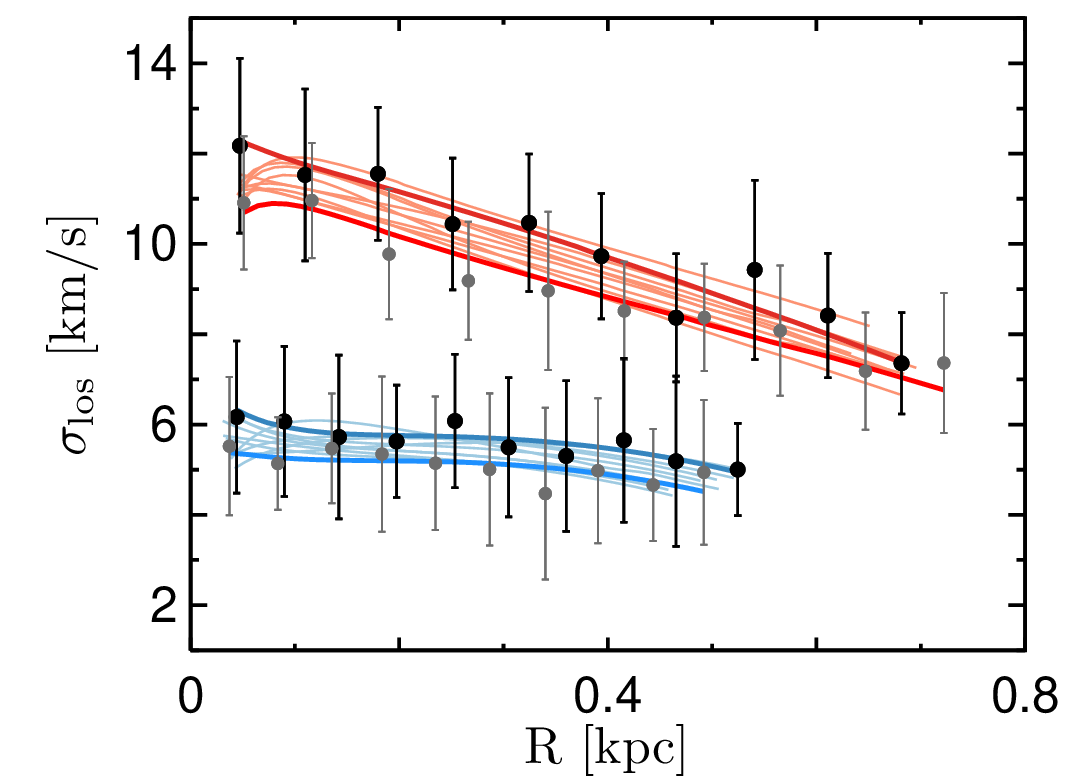}
\includegraphics[width = 0.49\textwidth]{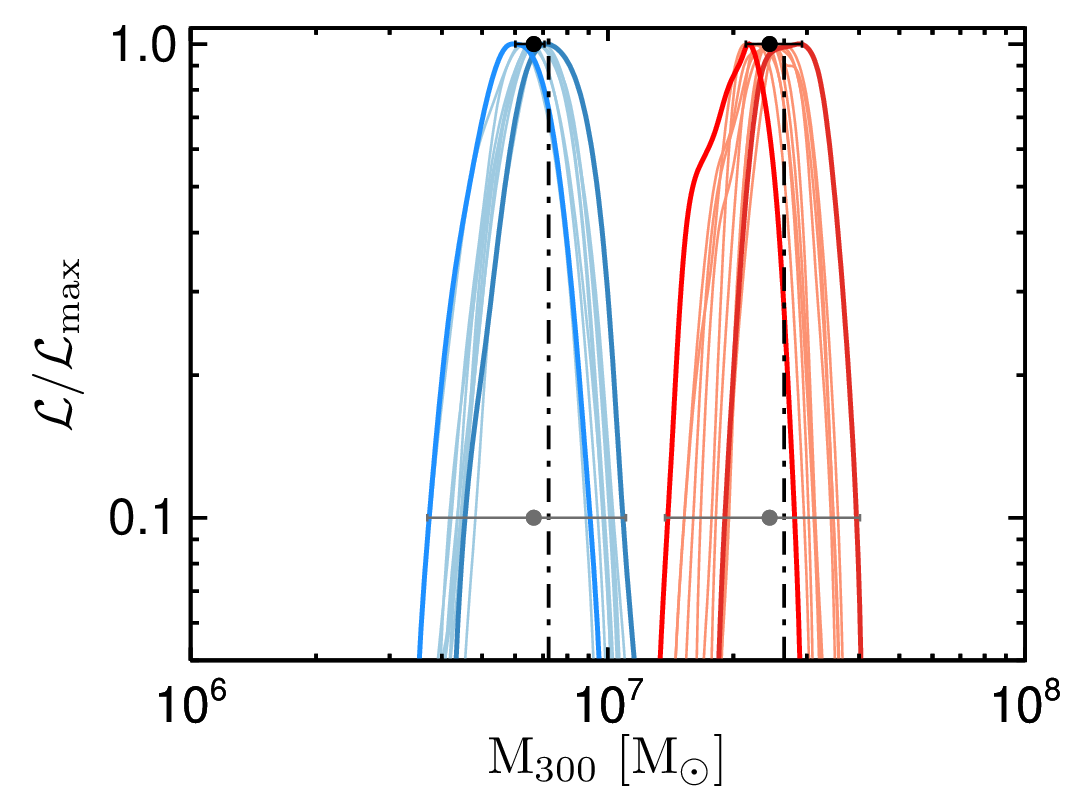}
\includegraphics[width = 0.49\textwidth]{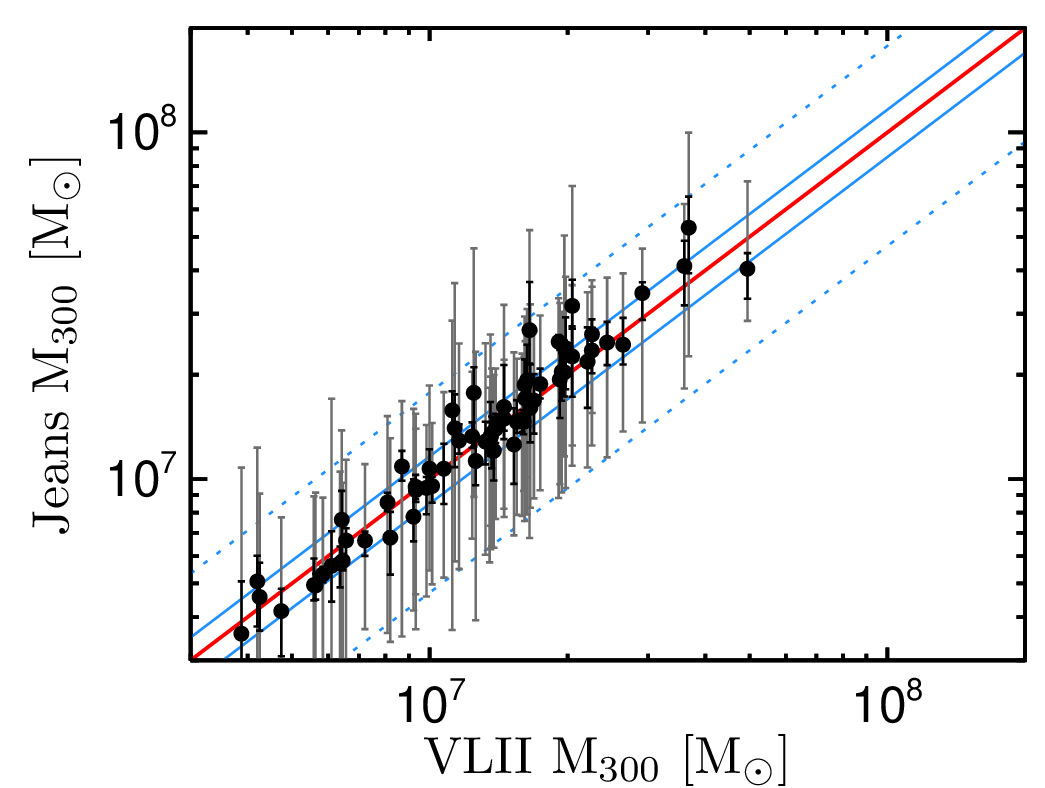}
\includegraphics[width = 0.49\textwidth]{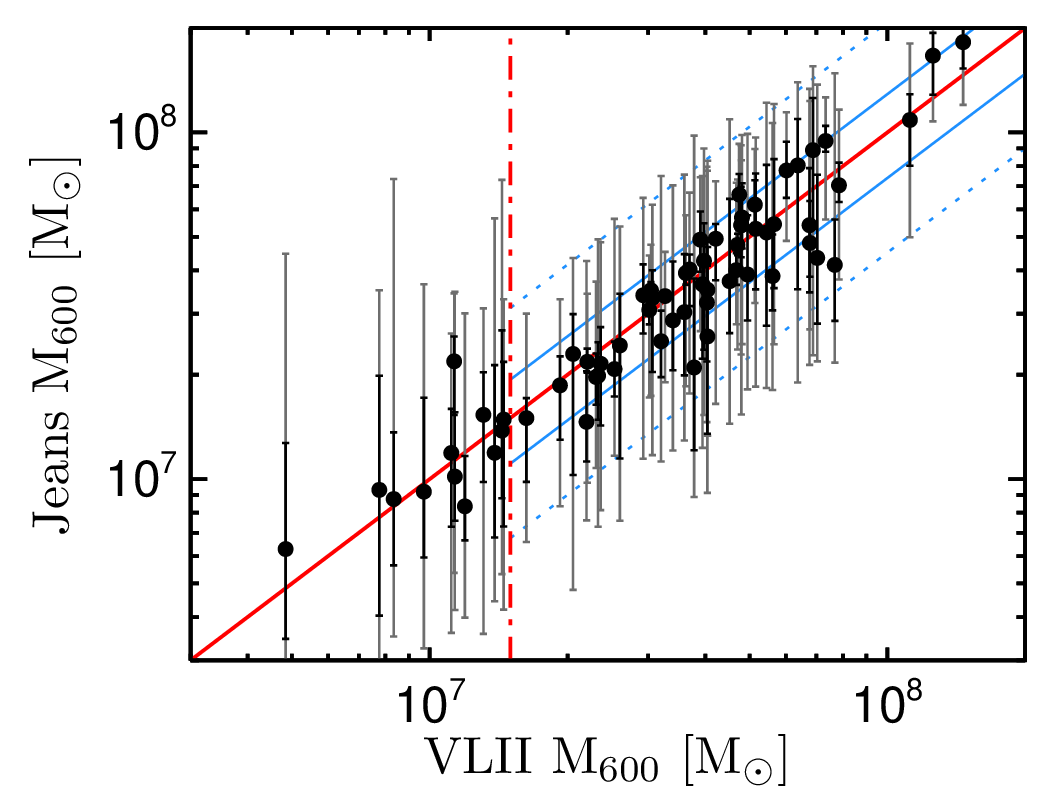}
\caption{{\it Left top panel:} Sample line-of-sight velocity dispersion profiles for two VLII dSph, of masses $M_{300} = 7.22 \times 10^6\,\msun$ and $M_{300} = 2.65 \times 10^7\,\msun$. Best fit Jeans solutions for $\sigma_{\rm los}(R)$ plotted in red and blue. Fits highlighted in darker colors correspond to the points with error bars and the envelope 10$\%$ likelihood functions shown in the right top panel. The average best fit estimates of $M_{300}$ in this case are $6.65 \times 10^6\,\msun$ and $2.44 \times 10^7\,\msun$ respectively.
{\it Right top panel:} Likelihood functions of $M_{300}$ for the two satellites in the left panel. The 11 realizations span a range in the best fit values around their mean (black dots with error bars) as well as a range in uncertainty where the likelihood has fallen to 10$\%$ of the best fit value (grey error bars). The vertical dot-dash lines indicate the true VLII values of $M_{300}$.
{\it Left bottom panel:} Total mass within the inner 300 pc, $M_{300}$ as determined through the Jeans equation in comparison with its true value from VLII. Mean values of 11 individual measurements are plotted with black points. Black error bars span the range between best fit values, while the grey error bars span the full range of values where the likelihood of the fit has fallen to 10$\%$ of its peak value. {\it Solid red line:} Unity correspondence. {\it Solid blue lines:} Average over the full $M_{300}$ mass range of the spread among the best-fit measurements. {\it Dotted blue lines:} Average over the full $M_{300}$ mass range of the spread among measurements where the likelihood is 10$\%$ of its peak.
{\it Right bottom panel:} Total mass within the inner 600 pc, $M_{600}$ as determined through the Jeans equation in comparison with its true value from VLII. Labels as in the $M_{300}$ panel on the left. Systems with $M_{600} \lesssim 1.5 \times 10^7\,\msun$ (left of the red dash-dot line) do not have kinematic data out to $r = 600$ pc, which results in a larger scatter in their estimated $M_{600}$ values.}
\label{fig8}
\vspace{+0.3cm}
\end{figure*}

Sample line-of-sight velocity dispersion profiles for two VLII simulated satellites and their assigned ``observational'' errors are shown in the top left panel of Figure~\ref{fig8}. The two satellites have masses of $M_{300} = 7.22 \times 10^6\,\msun$ and $M_{300} = 2.65 \times 10^7\,\msun$, respectively. For each satellite, we plot all 11 best fit $\sigma_{\rm los}(R)$ profiles. The two sets of data points (in black and grey) correspond to the kinematic profile fits highlighted in darker colors, which were chosen to the span the range of $M_{300}$ for which the fit's likelihood falls to 10$\%$ of its peak value (see the right top panel of Fig.~\ref{fig8}). The average $M_{300}$ values inferred from these 11 fits are $6.65 \times 10^6\,\msun$ and $2.44 \times 10^7\,\msun$ respectively, within 8$\%$ of the true values.

To get the above estimates, we fit a model to each ``observed'' kinematic profile using a Monte Carlo Markov Chain (MCMC). A theoretical velocity dispersion profile $\sigma_{\rm los,th}(R)$ is computed for each parameter set in step 2. and the likelihood of the fit is evaluated as:
\begin{equation}
\mathcal{L} = \prod_{i} \frac{1}{\sqrt{2\pi\sigma_i^2}}\textrm{exp}\left[-\frac{1}{2}\frac{(\sigma_{\rm los}(R_i)-\sigma_{\rm los,th}(R_i))^2}{\sigma_i^2}\right]
\label{eq:like}
\end{equation}
where $\sigma_{\rm los}(R_i)$ is the line-of-sight velocity dispersion measured at radius $R_i$ and $\sigma_i$ is its associated error. Each MCMC chain is initialized to maximize the likelihood $\mathcal{L}$ on a coarse model grid that spans the full parameter space. It is then left to explore the parameter space following the Metropolis-Hastings algorithm. Two concurrent chains are run on each data set and the MCMC is stopped after 10,000 attempts of each chain. This number is large enough to provide satisfactory fits to the kinematic data \citep{wal09}. The 11 resulting likelihood functions for $M_{300}$ of the two satellites above are shown in the right top panel in Figure~\ref{fig8}.

Once the best-fit model for each data set has been found, it provides the full mass density profile that can be integrated to any distance from the dSph center, with 300 pc or 600 pc being of particular interest. There are 11 likelihood distributions for the estimates of $M_{300}$ and $M_{600}$ of each VLII satellite, and these can be compared to the true values. The results are shown in the bottom panels of Figure~\ref{fig8}, in which we plot the average of the 11 best mass estimates as a function of the true dark matter mass. The nature of the MCMC data allows us to make two estimates for the error in each case. The smaller error bars (in black) in Figure~\ref{fig8} reflect the span of the 11 best fit values for each satellite. The larger errors (in grey) span the full range in values where the likelihood of the fit has fallen to 10$\%$ of its maximum value. For the particular two satellites in the top panels of Figure~\ref{fig8}, we thus have the following mass estimates: $M_{300} = 6.65 ^{+0.41 +4.40}_{-0.64 -2.95} \times 10^6 M_{\odot} $ for the lower-mass satellite and $M_{300} = 2.44 ^{+0.48 +1.48}_{-0.30 -1.07} \times 10^7 M_{\odot} $ for the higher-mass one, where the first set of errors corresponds to the best fit value range and the second set of errors corresponds to the 10$\%$ likelihood value range.

We infer that the spherical Jeans analysis is a robust method to estimate the central masses of dSph satellite galaxies. The mean best fit $M_{300}$ values determined through the Jeans analysis are on average within 12$\%$ of the true VLII values. The normalized scatter among the 11 best fits about the mean value is $^{+0.16}_{-0.15}$ on average and is shown as solid lines in the bottom panels of Figure~\ref{fig8}. The normalized full range scatter about the mean where the likelihood of the fit is greater than 10$\%$ of its peak value is $^{+0.78}_{-0.53}$ on average (dashed lines). No significant difference to these results arises when just the galactocentric observer's estimates are considered: the best fit galactocentric estimate of $M_{300}$ across the full mass range is on average within 16$\%$ of the true VLII value.  

When evaluating $M_{600}$, we exclude systems that have $M_{\rm 600} \lesssim 1.5 \times 10^7\,\msun$. It is apparent in the bottom right panel of Figure~\ref{fig8} that these 
exhibit a larger scatter than the systems with $M_{600} \ge 1.5 \times 10^7\,\msun$. The smallest dSphs host stellar systems smaller than 0.6 kpc in radial extent. The estimate of $M_{600}$ for them is therefore an extrapolation to a region in parameter space where tracers are simply not present. Among the dSphs with kinematic data available all the way to 
$r = 0.6$ kpc, the mean best fit $M_{600}$ value determined with the Jeans analysis is on average within 17$\%$ of the true VLII value. The normalized scatter among the 11 best fits about the mean value is $^{+0.29}_{-0.26}$ on average. The normalized full range scatter about the mean where the likelihood of the fit is greater than 10$\%$ of its peak value is $^{+1.08}_{-0.55}$ on average. The VLII galactocentric observer would report a $M_{600}$ value within 23$\%$ of the true VLII value.

\begin{figure*}
\centering
\includegraphics[width = 0.49\textwidth]{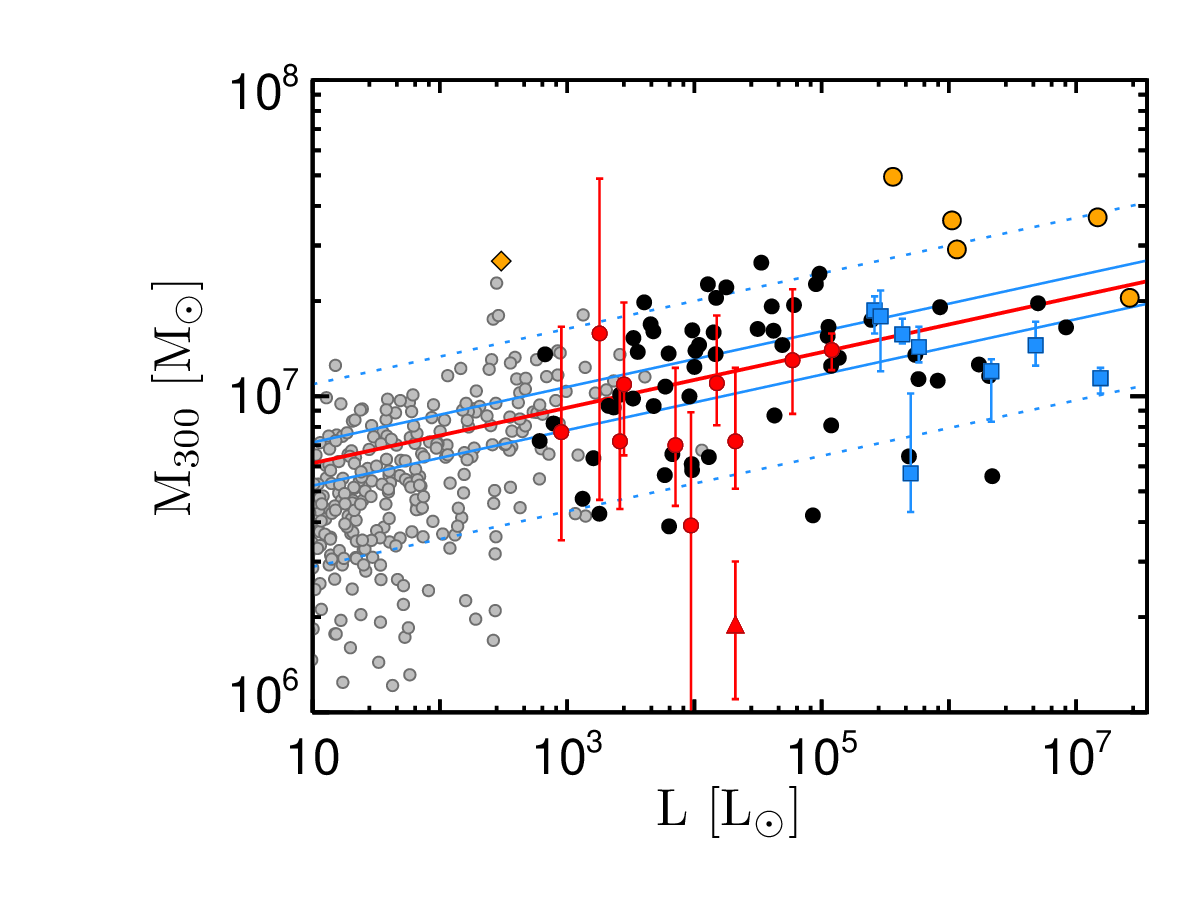}
\includegraphics[width = 0.49\textwidth]{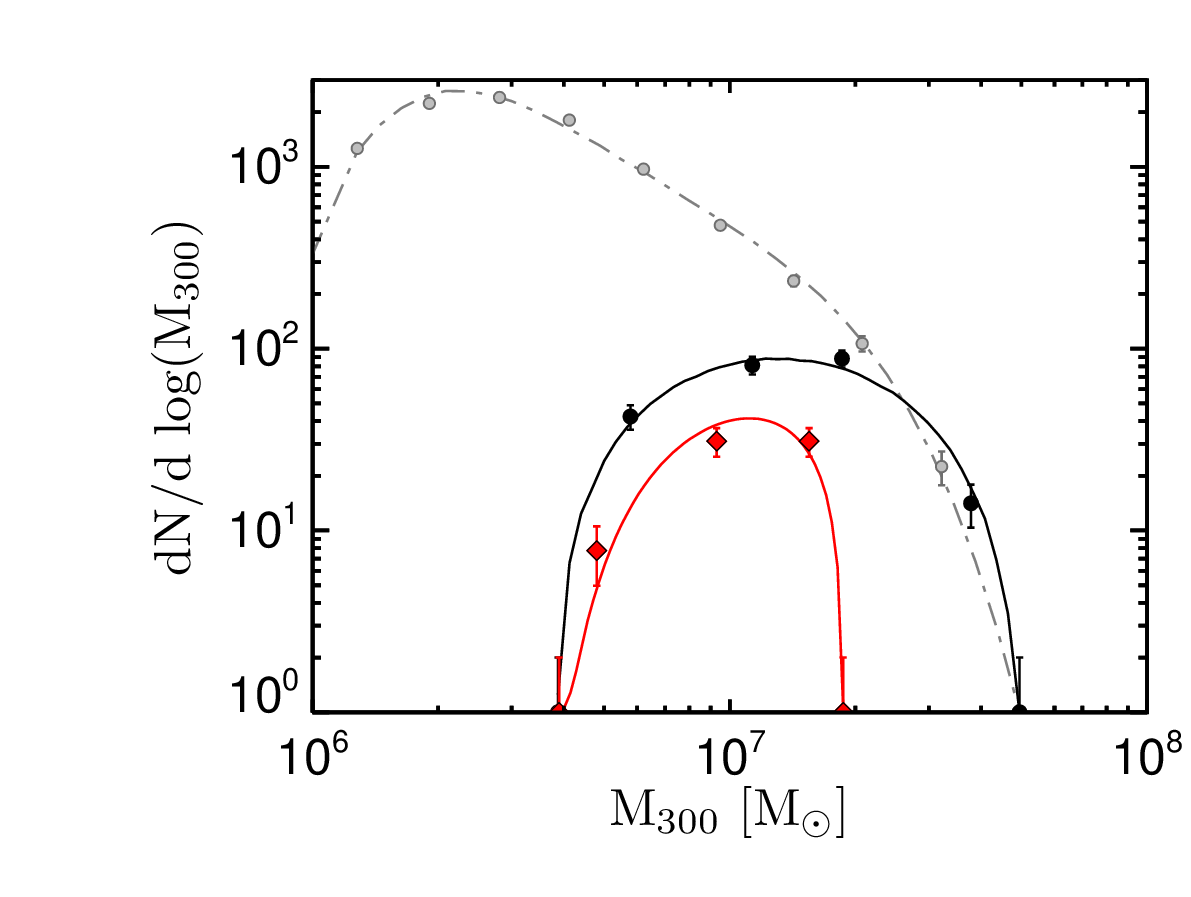}
\caption{
{\it Left panel:} Total mass within the central 300 pc, $M_{300}$ as a function of total luminosity for dwarf satellites in VLII  visible by an all-sky SDSS-like survey (black dots). Currently undetectable VLII dwarfs are shown in grey. A linear best fit to the data with slope 0.088 is shown with a red solid line. Data points with error bars are from \citet{str08}, indicating a ``common mass scale''. These error bars correspond to the points where the likelihood is 60.6$\%$ of its peak value on either side of the best estimate. For comparison, we plot our average error estimates in $M_{300}$ from Figure \ref{fig8} about the best fit ({\it solid red line}): {\it solid blue lines:} full spread in best-fit estimates; {\it dotted blue lines:} full spread in values where the likelihood is 10$\%$ of its peak. The ``classical'' dSphs ($L \ge 2 \times 10^5\,\lsun$) are shown as blue squares and the ``ultra-faint'' dSphs as red circles. The improved mass for the Hercules dwarf from \citet{ade09} is shown as a red upward pointing triangle. The six VLII subhalos identified by \citet{boy11} as too dense to host any of the known MW dSphs are plotted in orange as five circles (if detectable) and a single diamond (currently undetectable).
{\it Right panel:} $M_{\rm 300}$ mass function for VLII dSph satellites. {\it Red diamonds:} MW dSphs with masses taken from \citep{str08}. {\it Black dots:} VLII satellites detectable in SDSS. {\it Grey dots:} all VLII surviving subhalos that were tagged with stars, including those currently undetectable. The ``common mass scale'' is apparent in the currently detectable population.
}
\label{fig9}
\vspace{+0.3cm}
\end{figure*}

The range in $\sigma_{\rm los}(R)$ and $I_*(R)$ when measured from different viewpoints is due to the anisotropic velocity tensor of the dSph and any deviation from perfect sphericity of its shape. The mass estimate uncertainties among the best-fit values presented here should therefore be regarded as intrinsic, cosmologically-induced ones that will eventually be minimized when 3-D kinematic data is collected and better models of the nature of dSph satellites in high-resolution, fully cosmological hydrodynamical simulations are carried out.

In the left panel of Figure~\ref{fig9}, we plot the mass enclosed in the inner 300 pc versus total luminosity for all satellites that would be visible according to the SDSS detection limit. Overplotted 
are data from \citet{str08}, suggesting a ``common mass scale''. Observable VLII  dSphs exhibit a behavior similar to the MW dSph population, though a linear fit suggests a shallow but significant correlation between M$_{\rm 300}$ and the total luminosity: M$_{\rm 300} \propto L^{0.088 \pm 0.024}$, where the error on the slope reflects the 1$\sigma$ deviation from bootstrap re-sampling of the sample. This is steeper than the $L^{0.03 \pm 0.03}$ relation derived from observational data \citep{str08}. Lowering the derived $M_{300}$ values of some of the ultra-faint dSphs, as suggested by \citet{ade09} for the case of the Hercules dwarf, might be sufficient to reconcile the observations with the slope predicted from 
our model.

In our model, the nearly constant $M_{300}$ is explained by the combination of the steep adopted stellar mass-halo mass relation (eq. \ref{eq:mstar}) and the 
SDSS detectability magnitude limit. The remaining $\sim 2,000$ VLII surviving subhalos to which we assigned stars (about 200 of which are more luminous than $M_{\rm V}=0$) are just too faint to be detected by a SDSS-like survey. Some of them have $M_{300}$ significantly lower than $10^7$ M$_{\odot}$ (see the right panel of Figure \ref{fig9}).

\citet{boy11} has recently pointed out that state-of-the-art collisionless simulations of Galactic structure, both VLII and Aquarius, predict a population of massive 
subhalos that appear to be too dense (i.e. have too high a $V_{\rm max}$ for a given $r_{\rm vmax}$) to host any of the known MW dSph. While a failure of the CDM model 
is one interpretation of this result, another possibility is that star formation becomes stochastic below some mass, in which case the MW halo would harbor several 
massive but dark satellites. It is also quite plausible that baryonic feedback processes in the more massive subhalos redistribute some of their dark matter, thereby 
lowering their central density \citep[see e.g.][]{gov10}. For reference we have marked the six VLII subhalos that are too dense according to the \citet{boy11} 
analysis in the left panel of Figure~\ref{fig9}. Of these six, five have a luminosity greater than $M_V = -11$ and would have been detected as 
luminous dSph satellites. In our model, it is thus not possible for these overdense subhalos to have evaded detection because of an effectively low star formation 
efficiency resulting, for example, from their accretion or orbital history. As expected, these six halos have the highest values of $M_{300}$ for their luminosity 
among all VLII satellites. A reduction in $M_{300}$ by about a factor of two would bring them into agreement with the classical MW dSph sample. Our study of the standard Jeans analysis demonstrates that the `observed' M$_{300}$ values cannot be changed by that amount, i.e. that errors in the mass estimates are not responsible for the discrepancy between the models and the observations. Future, high-resolution hydrodynamical simulations of MW analogs that include an accurate treatment of baryonic processes and their effects on the central densities of luminous satellites will be crucial to the settling of this outstanding issue.

\section{Summary}
\label{sec:conclusion}

We have used a particle tagging technique to dynamically populate the $N$-body {\it Via Lactea II} high-resolution simulation with stars. The method is calibrated 
using the observed luminosity function of MW satellites, the concentration of their stellar populations, and the metallicity--luminosity correlation they exhibit, 
and self-consistently follows the accretion and disruption of progenitor dwarfs and the build-up of the stellar halo.  
The particle tagging technique makes full use of 27 simulation snapshots spanning the assembly history of the VLII main host between $z=27.54$ and $z=0$.   
A subset of the most bound dark matter particles in each of about 3,200 progenitor subhalos is tagged as stars at infall, when star particles acquire a 
stellar mass, an age, and a metallicity. Their subsequent evolution is purely photometric and kinematical in character, as the collisionless stellar populations age 
following population synthesis models, and are accreted and disrupted in a comological ``live host". Luminous particles linked to self-bound substructures 
in the $z=0$ snapshot constitute a surviving population of dSph satellites. Like the previous studies of \citet{bul05} and \citet{coo10}, we have shown that simple prescriptions for assigning stellar populations to 
collisionless particles are able to reproduce many properties of the MW dwarf satellite population, like velocity dispersions, sizes, brightness profiles, metallicities, 
and spatial distribution.
In agreement with previous results \citep[e.g.][]{kra10,kop09,mad08b}, a steep stellar mass-subhalo mass relation is required 
to fit the observed luminosity function of MW dSphs and offer an empirical solution to the missing satellite problem. This model predicts the existence of 
approximately 1,850 subhalos harboring ``extremely faint" satellites (with mass-to-light ratios $>5 \times 10^3$) lying 
beyond the SDSS detection threshold. Of these, about 20 are ``first galaxies", i.e. they formed a stellar mass above $10\,\msun$ before 
redshift 9. The ten most luminous satellites ($L>10^6\,\lsun$) in the simulation are hosted by subhalos with peak circular velocities today in the range $V_{\rm max}=10-40\,\kms$ 
that have shed between 80\% and 99\% of their dark mass after being accreted at redshifts $1.7<z<4.6$. The satellite maximum circular velocity $V_{\rm max}$ and 
stellar line-of-sight velocity dispersion $\sigma_{\rm los}$ today follow the relation $V_{\rm max}=2.2\sigma_{\rm los}$, i.e. stellar particles in satellites have 
smaller velocity dispersions than the dark matter. 
The ability to obtain simultaneous brightness and kinematic fits for dSphs like Sculptor and Carina is a confirmation of earlier searches for MW satellite counterparts in simulated halos such as the one by \citet{str10} and a success of our model. It is a prerequisite for studying another small-scale puzzle of galaxy formation in $\Lambda$CDM, namely the 
observed ``common mass scale'' of MW satellites, i.e. the fact that, despite spanning about 5 decades in luminosity, MW dSphs appear to contain a similar amount of 
dark mass, $M_{300}\sim 10^7\,\msun$, within a fixed aperture radius of 300 pc from their centers \citep{str08,wal09}.
We have applied a standard mass estimation algorithm based on Jeans modelling of the line-of-sight velocity dispersion 
profiles to the simulated VLII dwarf spheroidals, and tested the accuracy of this technique. We have shown that the mass enclosed within the inner 300 pc 
can be estimated accurately to within $20\%$ of the true value. The predicted $M_{300}$-luminosity relation for currently detectable simulated satellites 
is nearly flat. We do, however, predict a weak, but significant positive correlation for these objects: $M_{300}\propto L^{0.088 \pm 0.024}$. 
Our results are consistent with previous studies in which satellites observable by presently available wide-field surveys naturally tend to reside in subhalos 
that contain about $10^7\,\msun$ within their inner 300 pc. Future, even deeper surveys with the {\it Large Synoptic Survey Telescope} may be able to 
determine if an extension of this satellite galaxy population, at an even fainter magnitudes, indeed exists in the halo of the Milky Way. 
These experiments will cast light on the processes that govern galaxy formation on even smaller scales predicted in $\Lambda$CDM cosmology.
Ultimately, the prescriptions used in this work must be tested with hydrodynamical simulations that include baryonic physics. 
Simulations of the formation of late-type spiral galaxies in a cold dark matter ($\Lambda$CDM) universe have traditionally failed to yield realistic
candidates. Our group has recently reported a new cosmological hydrodynamic simulation of extreme dynamic range, ``Eris", in which a close analog of a Milky Way
disk galaxy arises naturally at the present epoch \citep{gue11}. With a mass resolution that is $\sim 20$ times lower than VLII,  
Eris includes radiative cooling, heating from a cosmic UV field and supernova explosions (blastwave feedback), a star formation recipe based on a high 
gas density threshold, and neglects any feedback from an active galactic nucleus. The study of the high-mass end of Eris' satellite population and the 
comparison with the results presented here will be the subject of a subsequent paper.

\section*{Acknowledgments}
This research was funded by NASA through grant NNX09AJ34G and by the NSF through grant AST-0908910. J.D. is supported by the Swiss National Science Foundation.
We thank J. Bullock, P. GuhaThakurta, and C. Rockosi  for many useful discussions and suggestions regarding this work, and Michael Obranovich for helping in coding the particle tagging technique. We also thank L. Strigari for providing sample kinematic data against which we tested our Jeans analysis code, and M. Walker for providing detailed observational data for MW dSph Carina and Sculptor.

\end{document}